\documentclass[useAMS,usenatbib,longnamesfirst,usegraphicx]{mn2e}

\usepackage{graphicx}

\def\apj{ApJ}

\def\mnras{MNRAS}
\def\nat{Nat}
\def\apjl{ApJ}
\def\aap{A\&A}
\def\aj{AJ}
\def\sovast{SvA}
\def\sci{Sci}
\def\araa{ARA\&A}
\def\apss{Ap\&SS}
\def\etal{et al.}

\title[The Vela and Geminga pulsars in the mid-infrared]{The Vela and Geminga pulsars in the mid-infrared}
\author[A.~A. Danilenko et al.]
{A.~A.~Danilenko,$^{1}$\thanks{E-mail: danila@astro.ioffe.ru}
 D.~A.~Zyuzin,$^{1,2}$ Yu.~A.~Shibanov$^{1,2}$ and S.~V.~Zharikov$^{3}$\\
$^{1}$Ioffe Physical Technical Institute, Politekhnicheskaya
26, St. Petersburg, 194021, Russia \\
$^{2}$St. Petersburg State Polytechnical Univ., Politekhnicheskaya 29, St. Petersburg, 195251, Russia \\
$^{3}$Observatorio Astron\'{o}mico Nacional SPM, Instituto de
Astronom\'{i}a, UNAM, Ensenada, BC, Mexico}

\begin{document}

\date{Accepted by MNRAS 2011 March 20. Received 2011 March 14; in original form 2010 October 8}

\pagerange{\pageref{firstpage}--\pageref{lastpage}} \pubyear{2011}

\maketitle

\label{firstpage}

\begin{abstract}
The Vela and Geminga pulsars are rotation powered neutron stars, which have been 
identified in various spectral domains, from the near-infrared to hard $\gamma$-rays. 
In the near-infrared they  exhibit tentative emission excesses, as compared to the optical 
range. To check whether these features are real,  we analysed archival mid-infrared 
broadband images obtained with the \textit{Spitzer} Space Telescope 
in the 3.6--160 $\mu$m range and compared them with the data in other spectral domains. 
In the 3.6 and 5.8 $\mu$m bands we detected at $\sim$ (4--5)$\sigma$ significance level 
a point-like object, that is  likely to be the counterpart of the Vela pulsar. Its position 
coincides with the pulsar  at $\la$ 0\farcs4 1$\sigma$-accuracy level. 
Combining the measured fluxes with the available multiwavelength spectrum 
of the pulsar shows a steep flux increase towards the infrared, confirming 
the reality of the near-infrared excess  reported early, 
and, hence, the reality of the suggested mid-infrared pulsar identification. Geminga is also 
identified, but only at a marginal 2$\sigma$ detection level in  one 3.6 $\mu$m band.     
This needs a farther confirmation by deeper observations, while   
the estimated flux is also compatible with the near-infrared Geminga excess.   
The detection of the infrared excess is in contrast to the Crab pulsar, where it is absent,  
but is similar to the two magnetars, 4U 0142+61 and 1E 2259+586, showing similar 
features. We discuss X-ray irradiated fall-back discs around the pulsars, unresolved pulsar nebula structures, and pulsar 
magnetospheres as possible origins of the excesses. We note 
also possible infrared signatures 
of an extended tail behind Geminga and  
of the Vela plerion radio lobes.
\end{abstract}


\section{Introduction}
\label{sec1}
The two nearby middle-aged pulsars, Vela (PSR B0833$-$45) and Geminga (PSR
J0633+1746), belong to a small group of rotation powered neutron stars (NSs),  which are 
identified in  various spectral domains outside the radio range.  Both pulsars have counterparts in the near-infrared 
(IR) and optical  \citep[e.g.,][ and references therein]{shib03,shib06,mign07}, in the ultraviolet (UV) 
\citep{Romani2005, Kar2005}, soft X-rays \citep[e.g., ][ and references therein]{pavlov2001, Kar2005, caraveo2004}, 
hard X-rays \citep{harding2002}, and   $\gamma$-rays \citep[][ and references therein]{Abdo2010v, Abdo2010g}. 
Emission mechanisms of pulsars are still poorly understood, and the presence of such multiwavelength  objects  is very 
useful for the progress in this field. 
       
In soft X-rays  and  extremal ultraviolet (EUV) the emission from the Vela and Geminga pulsars  is dominated by 
the thermal blackbody (BB) like radiation believed to be originated from cooling surfaces of the NSs, and, possible, 
from their hot polar caps. 
In the rest ranges the spectra are described by a power-law (PL)  with 
different spectral indices in different spectral domains.  
This  emission is  nonthermal and generated by relativistic  
particles accelerated in magnetospheres of the rapidly rotating 
and strongly magnetized NSs.

In the optical  the nonthermal spectra of both pulsars are significantly 
flatter  than  in soft X-rays,  and the optical fluxes  are considerably below the long 
wavelength extrapolations  of their nonthermal X-ray spectra. This suggests  a spectral break  
between the two ranges. Similar spectral picture is observed for the young Crab pulsar.   
In the near-IR the Vela and Geminga pulsars exhibit  apparent flux excesses, 
as compared to the optical range \citep{shib03,shib06}. This is in contrast to  the Crab, whose near and 
mid-IR fluxes are compatible with the spectral extrapolation from the optical range \citep{sol09,tem09}.  
The near-IR excesses appear to increase with the wavelength, and mid-IR  observations  
can be useful to confirm  or discard their reality. 
\begin{table}
\caption{\textit{Spitzer} observations of the Vela and Geminga pulsars.
}
\begin{tabular}{ccccc}
\hline\hline
AOR key&  Dev.  & Obs. date,  & Channels,      &    Int. time$^a$,                     \\
      &         & yy/mm/dd              & $\mu$m           &    s      \\
\hline
                         \multicolumn{5}{c}{the Vela pulsar}                                             \\
 \hline
11374848$^{b}$  & IRAC  & 04/12/17            & 3.6, 5.8, &    72          \\
 -  &   - & - &     4.5, 8.0 &      72       \\
11375104$^{b}$  & MIPS  & 05/03/07               & 24  &    92     \\
    -                            &     -        &             -                       &  70                &   335     \\
    -                           &     -        &             -                       &  160               &   105     \\
11542784$^{c}$ &  IRAC  & 04/12/18               & 3.6, 5.8 &  9400       \\
 \hline
                                         \multicolumn{5}{c}{the Geminga pulsar}                          \\
\hline
12540928$^{d}$    & MIPS  &  05/04/04  &  24               &  501                      \\
12543232$^{d}$  & IRAC  &  05/03/26  &  3.6, 5.8,      &  270                      \\
 - & -  & - &   4.5, 8.0      &            270            \\
19037696$^{e}$ &  IRAC  &  06/10/30      & 3.6,  5.8   &    14700                      \\
19037952$^{e}$ &   IRAC &  06/11/22          &  3.6,  5.8        &   14700             \\
\hline
\end{tabular}
\label{t:obs}
\begin{tabular}{ll}
$^a$~The integration time shown for the IRAC is the same for & \\
the each   channel in a given AOR & \\  
$^b$~PI, Patrick O. Slane, Program ID = 3647   & \\
$^c$~PI, Divas Sanwal, Program ID = 3696   & \\
$^d$~PI, Michael Jura, Program ID = 23   & \\
$^e$~PI, Deepto Chakrabarty, Program ID = 30822   & \\
\end{tabular}
\end{table}

Besides the Crab pulsar, there are only two isolated NSs detected  in the mid-IR.  
Both  are  anomalous X-ray pulsars (AXPs),  4U 0142+61 and   1E 2259+586.   
A strong mid-IR flux excess  was observed for the first one  
\citep{Wang2006}, and similar, but less evident  feature was found for the second   \citep{Kaplan2009}. 
The excesses can be interpreted   as   signatures of the so-called fall-back discs which can be formed around  NSs 
shortly after the supernova (SN) explosions. Such  discs can be  progenitors of pulsar planetary systems, one of which 
was discovered  around  PSR B1257+12  \citep{Wols1994,konacki03}.  However, the association of the excesses 
with the intrinsic pulsar emission still can  not be  excluded.   
  
In addition to their multiwavelength emission, the Vela and Geminga pulsars power extended pulsar wind nebulae 
(PWNe) most clearly visible in soft X-rays \citep{Helfand2001, pavlov2001a, caraveo2003, deluca2006, pavlov2006, 
pavlov2010}. The PWN morphologies  are different.   
The younger Vela ($\sim$ 10$^4$ yr old) is associated with the supernova remnant 
(SNR)  G263.9$-$3.3 and, as the Crab pulsar, forms  a torus-like inner PWN with polar jets. 
The PWN of the older Geminga ($\sim$ 3.4$\times$10$^{5}$ yr) has a tail and/or a bow-shock morphology.  
At larger spatial scales the Vela PWN has a cocoon, or a plerion, structure. 
It is  mainly  elongated southwards 
the pulsar, that is roughly orthogonal to the orientation of the inner PWN symmetry axis \citep{maog1995, Dod2003a}.  
The cocoon has been detected 
in the GeV range with \textit{Fermi} \citep{Abdo2010vpwn} and the AGILE \citep{pell2010}, 
and in the TeV range with the HESS \citep{ahar2006}. It is speculated, that the high energy source may be a 'relict' PWN 
distorted by the passage of the remnant reverse shock in inhomogeneous medium \citep{blondin2001}.  
Geminga is not associated with any SNR,  although,   it is discussed \citep{salvati2010}, 
that a dissolved and not visible Geminga SNR   could be a plausible source  of a small scale excess  
in multi-TeV cosmic ray protons  arrival direction (spot A)  discovered  with Milagro \citep{Abdo2008}.  

Both PWNe   have no obvious optical counterparts.  Some faint 
extended features were detected in the near-IR with the VLT in the 1\farcs5--3\farcs1 
vicinity of the Vela pulsar \citep{shib03}.   They are 
projected on and/or aligned with the south-east jet and the inner arc of its X-ray PWN. 
There is also a faint optical nebulosity with a size of few arcsec around the Geminga pulsar 
detected in  a deepest up to date optical image obtained in the \textit{I} band with the Subaru 
telescope \citep{shib06}. The nebula is extended perpendicular of the pulsar proper motion, and, probably, is the
brightest inner part of the bow-shock structure which is seen in X-rays.
However, the PWN origin of those features still needs to be confirmed.
 
Motivated by  further studies of the IR flux excesses in the emission of both pulsars and by searching 
for their PWN IR-counterparts, in this paper we analyse unpublished archival mid-IR  images 
of the pulsar fields  obtained with the \textit{Spitzer} Space Telescope. We report on finding likely 
mid-IR counterparts of the pulsars and possible features of their PWNe. The mid-IR observations are 
described in Sect. 2, the results are presented and discussed in Sect. 3 and 4.

\section{Observations and data analysis}
\label{sec2}
\subsection{\textit{Spitzer} data}
The Vela and Geminga pulsar fields have been imaged several times with the
\textit{Spitzer} telescope using the Infrared Array Camera (IRAC)
and the Multiband Imaging Photometer for \textit{Spitzer} (MIPS)\footnote{see http://ssc.spitzer.caltech.edu}.  
Various instrument channels have been used with the  effective wavelengths of 3.6, 4.5, 5.8, 8, 24, 70, 
and 160 $\mu$m.  The information on the respective Astronomical Observation Requests (AORs) 
is presented in Table \ref{t:obs}. We retrieved the data from the \textit{Spitzer} archive.  
We used the IRAC  post-BCD (Best Calibrated Data)  mosaic images with an effective image scale 
of 0\farcs6 per pixel and the field of view (FOV) of 7\farcm5$\times$7\farcm5. They have been  generated with the \textit{Spitzer} pipeline 
 procession version of S18.7.0. 
We also regenerated the MIPS post-BCD calibrated mosaic images from respective BCD data using 
the MOPEX\footnote{see http://ssc.spitzer.caltech.edu/postbcd/mopex.html} tool v18.3.1. The resulting image scales 
are 2\farcs4 per pixel for the 24 $\mu$m MIPS channel with the FOV of 13\farcm3$\times$12\farcm8. 
After the pipeline drizzling of multiple individual exposures  the effective pixel scale 
for the IRAC  mosaics is twice smaller than the nominal one (1\farcs2 per pixel), 
while for the MIPS it is consistent  with the nominal pixel scale  of 2\farcs4.  

First inspections of the archival data showed that  a point-like counterpart  candidate of the Vela pulsar 
is   detected at $\ga$ 4$\sigma$ significance level in the  3.6 and 5.8 $\mu$m images of the 
AOR 11542784, where   
the integration time   was maximal among the other AORs for this target (cf. Table \ref{t:obs}). 
The respective mosaic coverage maps show that the real integration times of the pulsar neighbourhood 
located in the centre of the mosaic   images are only 
1--3 \% smaller  than the AOR nominal archival integration time shown in Table~\ref{t:obs}.    
At the shorter   AOR 11374848  the  source  is probably marginally detected  at  3.6, 4.5, and 5.8 $\mu$m,  
and it is not detected at 8 $\mu$m.  A bright pixel at the source position is seen  
in the 24  $\mu$m image  (AOR 11375104), while at longer wavelengths no signatures 
of the pulsar  are detected.  

A likely counterpart  of the Geminga pulsar is marginally detected at  $\sim$ 2$\sigma$ significance level in the 3.6  $\mu$m  
images of the long integration time AORs 19037696 and  19037952.  
These AORs  differ by a zodiacal background level, that was $\sim$ 30 \% smaller in  the second AOR. 
At  5.8 $\mu$m  the candidate is not resolved  even  in  the image  combined from the data of the two  
AORs. However,  we found some sign of it in the 4.5 $\mu$m image of the shorter AOR 12543232.   
In other \textit{Spitzer} bands and shorter AORs we have not found any signature of the pulsar.  

To check that the detected counterpart candidates are not the detector artefacts or  results 
of a poor cosmic ray rejection, we analysed the respective badpixel masks, provided by the archive together 
with the data, and regenerated the  mosaic images from the BCD data using  the MOPEX tool v18.3.1.  
We used    the nominal  IRAC pixel scale and a half of it, as has been done in the pipeline processed images discussed above.  
For   the Vela AORs we obtained the similar results, confirming the reality of the counterpart candidate 
detection in both IRAC bands of the AOR 11542784 and the presence of a  bright pixel at the source position in  
the 24  $\mu$m image  of the AOR 11375104.  
There is a  bright star located in 0\farcm5 north-east of the Vela pulsar. In 3.6 and 5.8 $\mu$m images it is oversaturated  
and produces  typical stray light artefacts.  However, the nearest one is in $\sim$ 8 arcsec  south-east of the pulsar position 
and cannot induce  a false source detection at least in a few arcsec region around the pulsar.   
 The marginal Geminga counterpart in 3.6 $\mu$m images of the AORs 19037696 and  19037952 was found 
to be present at the same significance level in the nominal and the half pixel scale regenerated mosaics. 
However, at the 4.5  $\mu$m it  disappears in the nominal pixel scale mosaic. Its apparent presence 
in the half pixel scale  drizzled mosaic is likely to be an artificial background fluctuation  resulting from  an  
inappropriate use of the   0\farcs6 pixel scale  in the pipeline data procession for a shallow observation in this band, 
where only nine separate exposures have been obtained. Therefore, only an upper limit on the Geminga flux can be set in this band. 
Further investigation of the 3.6 $\mu$m images of  Geminga showed that  they 
contain some artefacts. These are so called 'muxbleed' and 'muxstripes', which are  due to the presence of 
two  bright stars in the FOV north-west of Geminga and  
which cannot be perfectly corrected by the reduction tool at given AOR setups\footnote{For details, see,~e.g.,~the~IRAC~Instrument~Handbook, 
http://ssc.spitzer.caltech.edu/irac/iracinstrumenthandbook/}. The respective stripes cross the whole mosaics from north to south approximately.  
However, the nearest one is located in $\sim$ 0\farcm5  west of the Geminga position  and cannot 
result in a false source detection, at least in   several tenths arcsec pulsar neighbourhoods. 

Thus, we can conclude, that the Vela counterpart candidate in the 3.6 and 5.8 $\mu$m images is definitely the real source. 
The Geminga candidate in 3.6 $\mu$m images is likely to be the real source as well, but it is much fainter and detected 
at a lower significance level. The source reality  is additionally supported by  
its independent detection at about the same significance level in the two AORs, which were obtained at different epochs  
and conditions.  
Below we mainly focus on the  data where our counterpart candidates have been checked to be the real mid-IR sources.

 \subsection{Astrometry}
The coordinates and proper motions of the Vela and Geminga pulsars  are known 
with a high accuracy based on 
the radio and optical observations \citep{caraveo2001, Dod2003, bign1993, fah07}.    
For correct positional identifications of the pulsars with the    sources  detected 
in the mid-IR, one has 
to be sure in  the \textit{Spitzer} pointing accuracy. 
To check this in the  IRAC images of both pulsars, we selected  about two dozens of  unsaturated 
isolated astrometric standards from the USNO-B1 catalogue.   
Their catalogue  and image positional uncertainties for both 
coordinates  were $\la$ 0\farcs2 and $\la$ 0\farcs5, respectively.
IRAF  {\sl ccmap/cctran} tasks  were used to
find plate solutions. Formal rms uncertainties of the
astrometric fit were $\la$ 0\farcs16  with maximal  residuals of
$\la$ 0\farcs4 for both coordinates. After astrometric transformations the shifts 
between the original and
transformed images were $\la$ 0\farcs4 for the Vela  and $\la$ 0\farcs2 for 
the Geminga fields. This is less than the  nominal pixel
scale (1\farcs2) of the IRAC, ensuring the almost perfect pointing accuracy of the
\textit{Spitzer} observations. Combining  all uncertainties,  a conservative
estimate of the 1$\sigma$ referencing uncertainty is  $\la$ 0\farcs4  in both coordinates for all
IRAC images in question. This is comparable with  astrometric 
uncertainties of   available \textit{Chandra}/ACIS and HRC images of   the same fields in X-rays,  
but about twice as poor as for  optical and near-IR  images  obtained with large 
ground based telescopes 
and the \textit{HST}.  Nevertheless, this  allows us to identify positionally on a subarcsecond 
accuracy level the mid-IR, near-IR, 
optical and X-rays  objects in the respective pulsar frames, which have been obtained 
with significantly different spatial resolutions and pixel scales.   
\section{Results}
 \label{sec3}
 \subsection{Morphology of the Vela field} 
 \subsubsection{Identification of the Vela pulsar} 
\begin{figure*}
 \setlength{\unitlength}{1mm}
\resizebox{12.cm}{!}{
\begin{picture}(120,128)(1,0)
\put (-27,65) {\includegraphics[scale=0.54, clip]{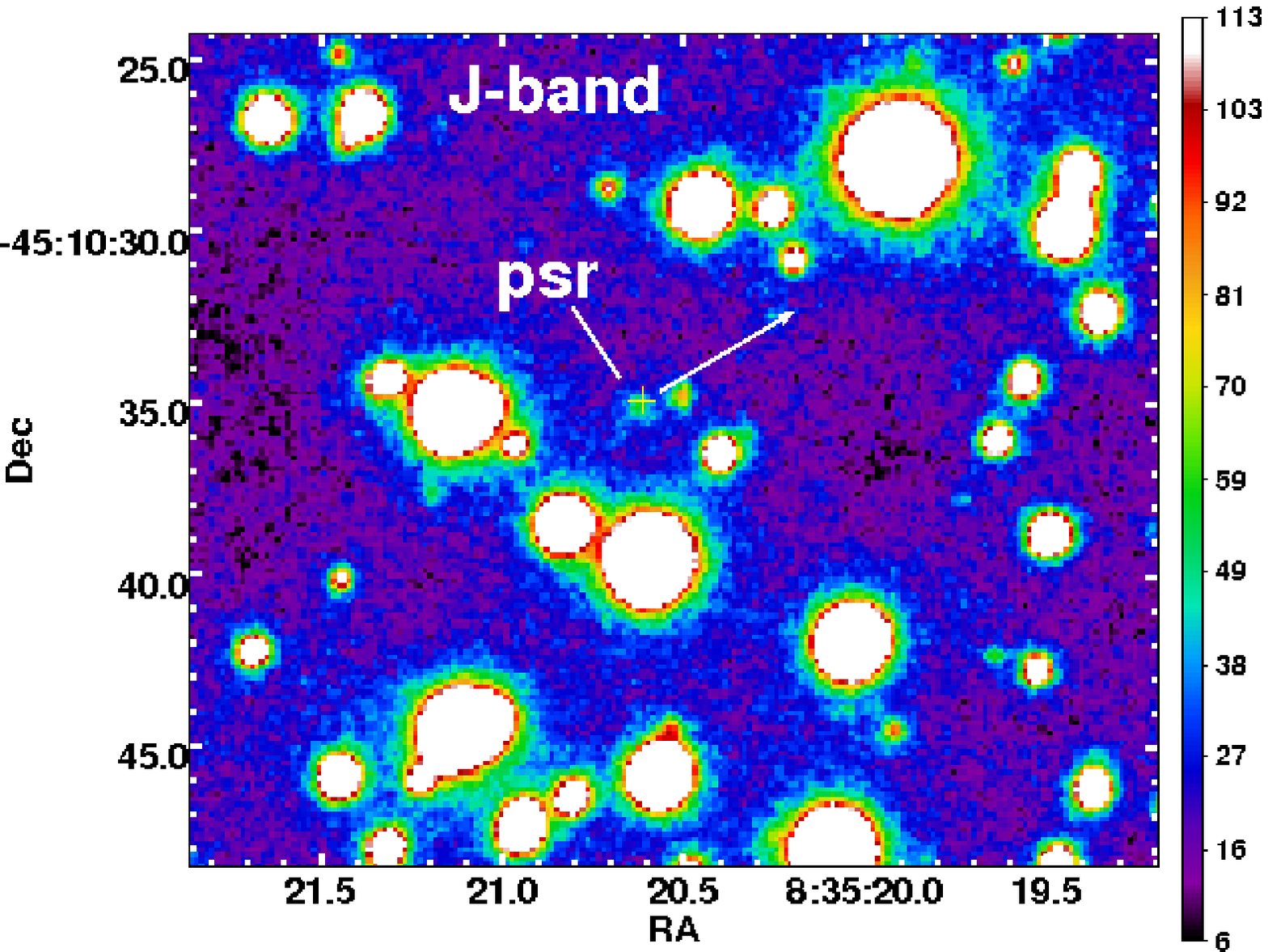}} 
\put (62,65)  {\includegraphics[scale=0.54, clip]{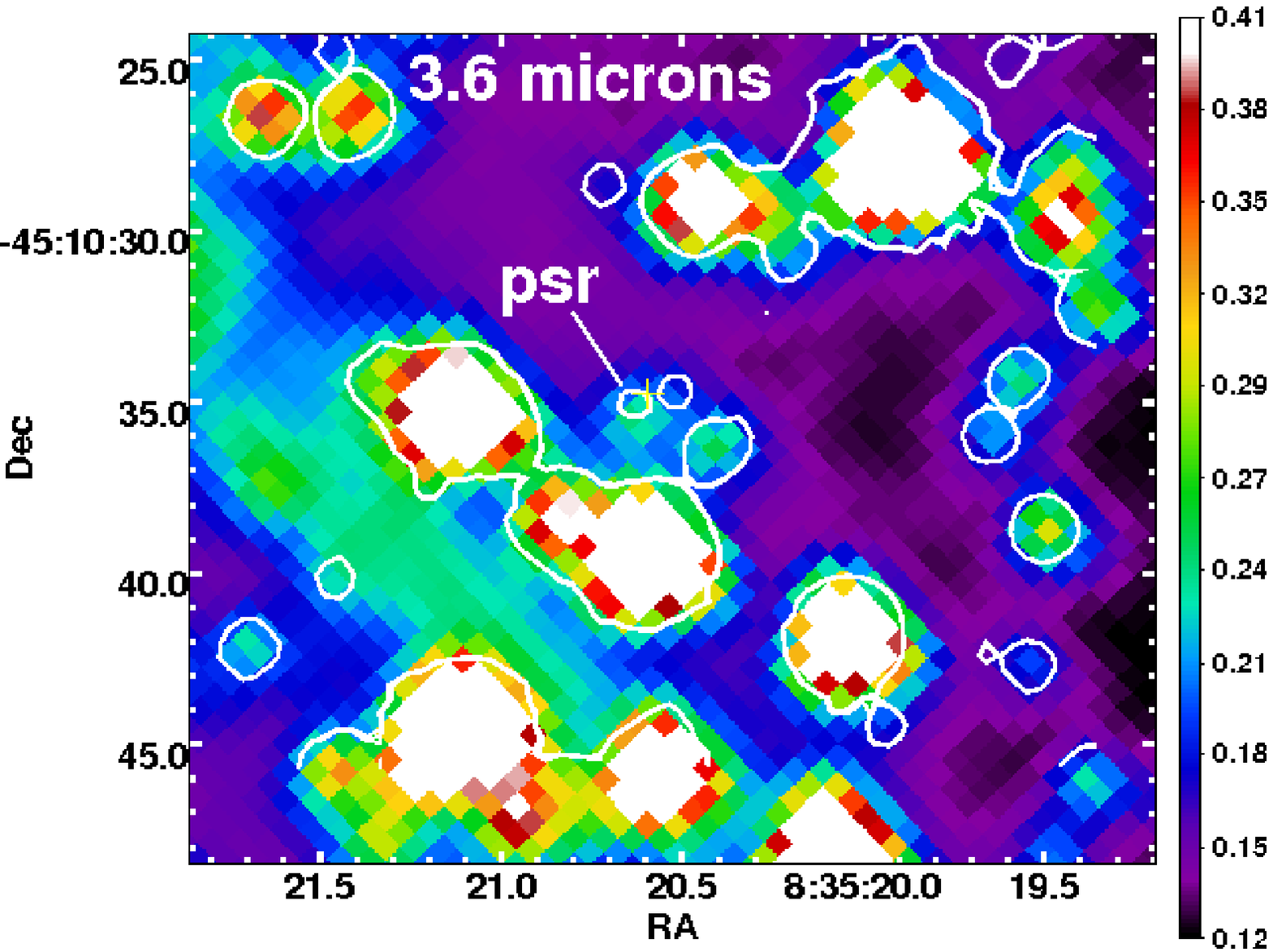}} 
\put (-27,0)  {\includegraphics[scale=0.54, clip]{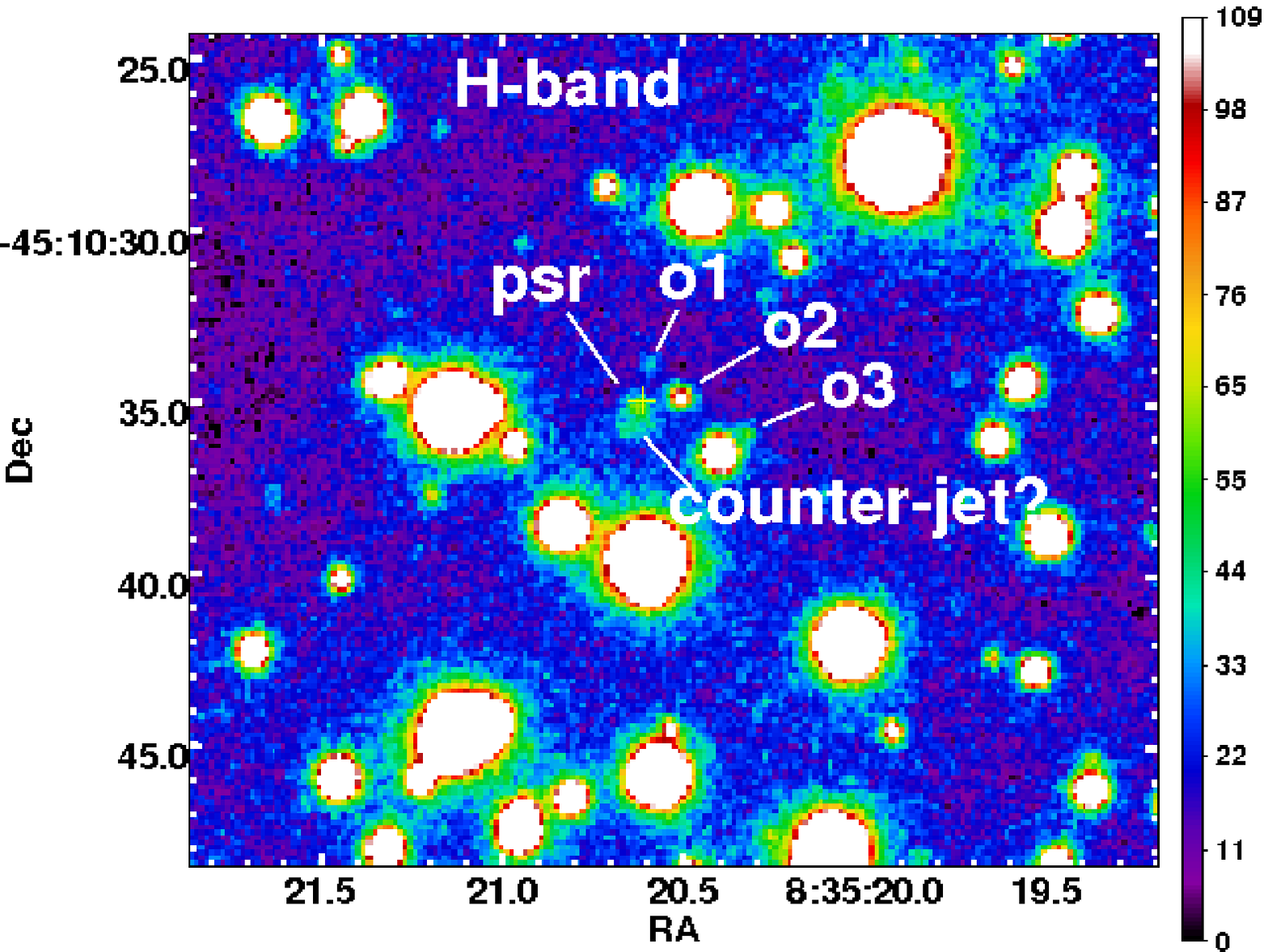}} 
\put (62,0)   {\includegraphics[scale=0.54, clip]{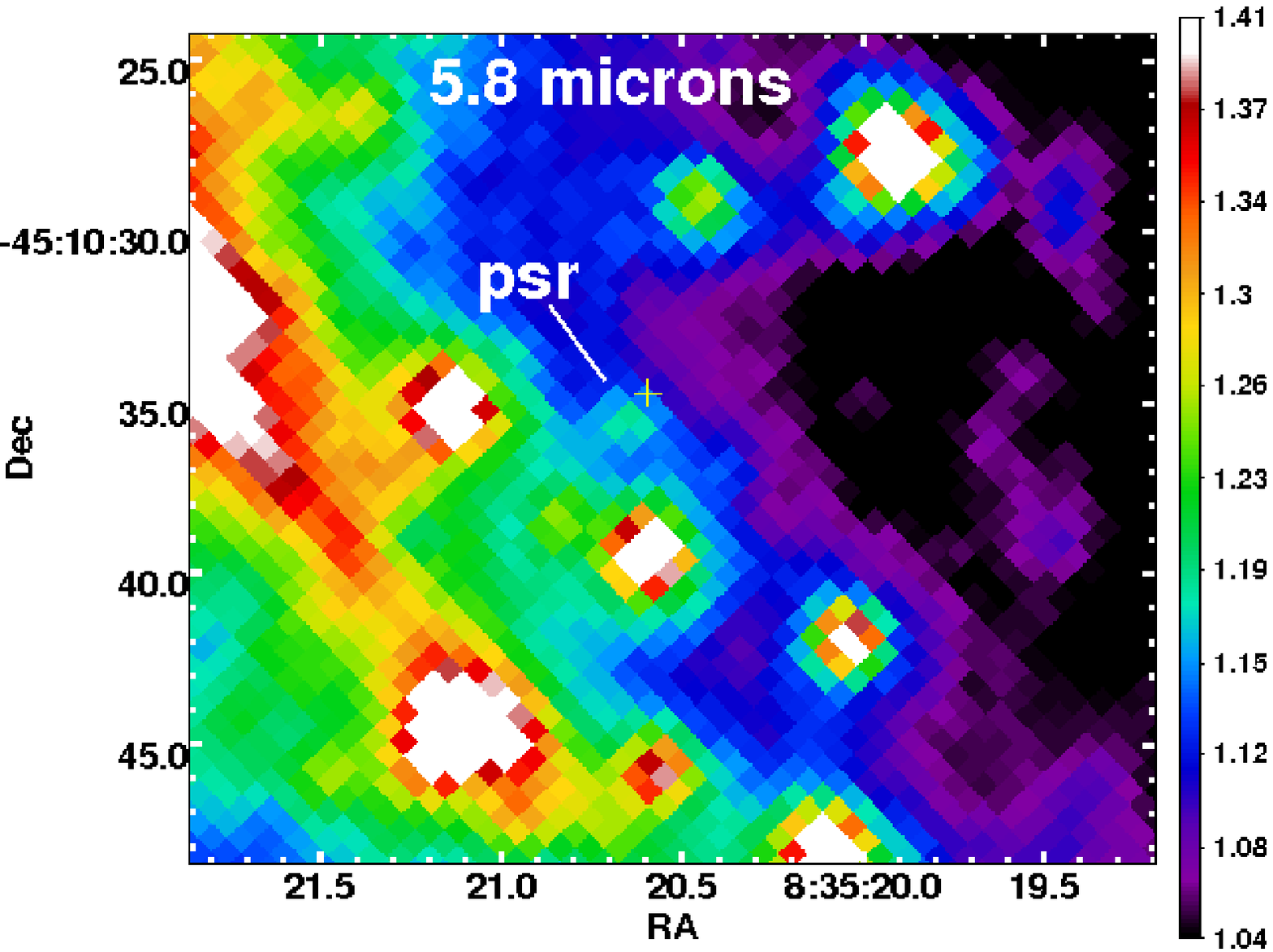}} 
 \end{picture}}
 \caption{  The  $\sim$ 20\arcsec$\times$25\arcsec~fragments of the infrared images 
 of the Vela field  obtained with the VLT/ISAAC  in the \textit{J} and \textit{H} bands ({\sl left column}), and  
 with the \textit{Spitzer}/IRAC (AOR 11542784) in 3.6  and 5.8 $\mu$m  bands ({\sl right column}). 
 The Vela pulsar counterpart is marked by a thick line and the arrow shows   
 the direction of its proper motion. Crosses  are the expected pulsar radio positions defined  separately for the VLT and \textit{Spitzer} observation 
 epochs.   A  knot-like object 'o1', an  extended 
 feature SE of the pulsar, possible associated with the X-ray PWN jet and counter-jet,  and two background stars 'o2' and 'o3'  
 are notified  in the \textit{H} band image. The contours in the 3.6 $\mu$m images are from the \textit{J} band  image. 
 Henceforth the image pixel value scale is linear.}
 \label{fig:vela-ima-1}
\end{figure*}

In Fig.~\ref{fig:vela-ima-1} we compare fragments of the 3.6 and 5.8 $\mu$m \textit{Spitzer} images 
containing the Vela pulsar  with the \textit{J} and \textit{H} band images of the same field obtained with
the VLT/ISAAC in  the period of December 2000 -- January 2001 \citep{shib03}. 
The point-like near-IR  pulsar  counterpart is  marked by a thick line, and it is 
clearly visible   
near the same position in the mid-IR. The measured coordinates of the counterpart 
candidate in the 3.6   $\mu$m image, where it is detected with a higher significance than at  
5.8 $\mu$m,  are   RA=08$^{h}$ 35$^{m}$ 20$^{s}$.635$\pm$0$^{s}$.042 and Dec=$-$45$^{\circ}$ 10\arcmin~35\farcs48$\pm$0\farcs45 (J2000). The errors 
account for $\approx$ 0\farcs12 object position uncertainties in  the image and 
the mid-IR astrometric 
referencing uncertainties estimated in Sect.~2.2. 
%
Within 1$\sigma$ the coordinates are  consistent with the position of 
the near-IR counterpart, 
RA=08$^{h}$ 35$^{m}$ 20$^{s}$.628$\pm$0$^{s}$.01 and Dec=$-$45$^{\circ}$ 10\arcmin~35\farcs16$\pm$0\farcs1  (J2000), which is 
defined with a higher accuracy due to a  better spatial resolution 
of  the VLT observations obtained with the pixel scale of 0\farcs147. 
The contours overlaid on the 3.6 $\mu$m image from the \textit{J} band image underline 
the positional consistency of the detected mid-IR candidate and the near-IR counterpart.
 They  show also that all background  near-IR sources, except of the two faint 
 near-IR point-like 
 objects 'o1' and 'o2' located in immediate vicinity north-west of the pulsar, are identified in the 
 mid-IR and vice versa. Hence, the appearance of a new mid-IR background 
 source at the pulsar position, not visible in the adjacent near-IR band,  
 is very improbable.  An apparent  emission region extended 
 from north-east to south-west  in $\sim$ 8 arcsec 
  south-east of the pulsar is the  stray light artefact  from a bright saturated star 
  located  in $\sim$ 0\farcm5 north-east of the presented frames.  
  This has been mentioned  in Sect. 2.1 and may somewhat increase     
  the backgrounds level near the pulsar position and decrease the significance 
  of the candidate detection.       
  
The pulsar proper motion is $\mu$ = 58$\pm$0.1 mas yr$^{-1}$ with a positional angle of 
301\fdg0$\pm$1\fdg8 \citep{Dod2003}. Hence, the pulsar shift during four years between 
the VLT and \textit{Spitzer} observation epochs is $\approx$ 0\farcs23, that is indistinguishable  
within the derived positional error budget. 
Attempting to improve the accuracy   
of the positional identification we have   performed the relative astrometry. 
The VLT \textit{H} band image with  a point-source   FWHM $\approx$ 0\farcs5   
was used as a reference for finding the plate solution for the 3.6   $\mu$m image   with    FWHM $\approx$ 1\farcs8.    
The IRAF { \sl geomap/geoxytran} tool with  a { \sl general} geometry option, allowing for 
the  image shifting,  rotation,  pixel scaling, and a skew,   was applied.
 A dozen of  unsaturated isolated stars,   
which are common for both frames and are located in  vicinity of the pulsar, were selected as reference points.   
The formal rms uncertainties of the plate solution fit were $\la$ 0\farcs09  with maximal  residuals of
$\la$ 0\farcs17 for both image coordinates. The  pixel scale ratio was  $\approx$ 0.245, that coincides  
with  the nominal one for the two images.  
  Using this solution we transformed the  3.6   $\mu$m image coordinates of the mid-IR counterpart candidate to 
the \textit{H} image.  As a result, we obtained that the candidate counterpart is apparently displaced
 by  0\farcs41$\pm$0\farcs21  to the south from the near-IR pulsar counterpart. 
 Similar displacement   can be noticed also from absolute astrometry results,  which have been used in   Fig.~\ref{fig:vela-ima-1}, where   
 the mid-IR candidate position   in the 3.6  $\mu$m image is 
also slightly shifted toward the south from the near-IR pulsar contour. 
In any case, the shift  significance is too small, $\la$ 2$\sigma$,  to consider it seriously as a real one.  
Nevertheless, this may be a signature of a systematic shift caused by  a faint extended 
feature located in the immediate vicinity southward of the pulsar and marked as a 'counter-jet' in the \textit{H} band panel of Fig.~\ref{fig:vela-ima-1}. 
The feature  is  likely associated with the pulsar PWN and  is clearly detected in the near-IR \citep{shib03}, 
but it cannot be resolved from the point source in the mid-IR due to a poor Spitzer spatial resolution. 
We shall discuss the implications of that  later in Sect. 4.1.2.  
Using the plate solution and absolute astrometry of the \textit{H} band image, we obtain the candidate coordinates, 
RA=08$^{h}$ 35$^{m}$ 20$^{s}$.626$\pm$0$^{s}$.025 and Dec=$-$45$^{\circ}$ 10\arcmin~35\farcs51$\pm$0\farcs26  (J2000), which are consistent  
with  those obtained 
from the direct  mid-IR absolute astrometry, while they  are formally about twice more accurate  than the latter ones.  
However, these improvements  do not still allow us to distinguish the counterpart  proper motion. 
The  new formal position error budget is still comparable with the proper  motion shift of the pulsar, 
and including the possible systematic shift discussed above increases the positional uncertainties to the values obtained 
early from the absolute astrometry.        

The expected radio coordinates of the pulsar at the mid-IR 
observation epoch are RA=08$^{h}$ 35$^{m}$ 20$^{s}$.588$\pm$0$^{s}$.038 and Dec=$-$45$^{\circ}$ 10\arcmin~34\farcs727$\pm$0\farcs4 (J2000).   
They are derived from   the most accurate VLBI  radio position and  proper motion measurements  
\citep{Dod2003}, and the errors  are dominated by the mid-IR astrometric referencing 
uncertainties. The displacement between the expected position and the mid-IR candidate
centre is $\sim$ 0\farcs9, or about (2--3)$\sigma$  of the coordinate uncertainty level. 
This can be resolved in Fig.~\ref{fig:vela-ima-1}, where  radio positions are marked by  crosses.  
The  direction of the offset approximately coincides with the pulsar proper motion positional angle. 
Considering available optical/near-IR images obtained early with 
the VLT and \textit{HST} and referenced with the USNO catalogue we got  similar results: 
the  expected radio positions  are  always marginally shifted by  about  
0\farcs2--0\farcs4 ahead of the counterpart positions approximately along  
the line of the pulsar proper motion (cf. Fig.~\ref{fig:vela-ima-1}).
The variation of the offset values  can be   
explained by different spatial resolution of the observations, 
while the same direction of the offsets   is  noticeable.
At the same time, the relative values, 
such as  the counterpart  motion and its shifts from one epoch to others, 
are compatible with the radio proper motion. Dodson et al. (2003) have discussed 
 that some differences between 
the radio and optical absolute astrometries  of the pulsar can be due 
to different  astrometric standards used in the two ranges, i.e., very accurately established  
extragalactic standards in the radio  and mainly Galactic  standards in the optical, 
where proper motions are sometimes poorly  known. Therefore,  the apparent offsets  
of the optical/IR counterpart from the radio position of the pulsar  are likely caused  
by  unknown systematic uncertainty of the optical astrometric standards for the Vela field. 
\begin{figure}
 \setlength{\unitlength}{1mm}
\resizebox{12.2cm}{!}{
\begin{picture}(120,102)(1,0)
\put (-0.8,51) {\includegraphics[scale=0.52, clip]{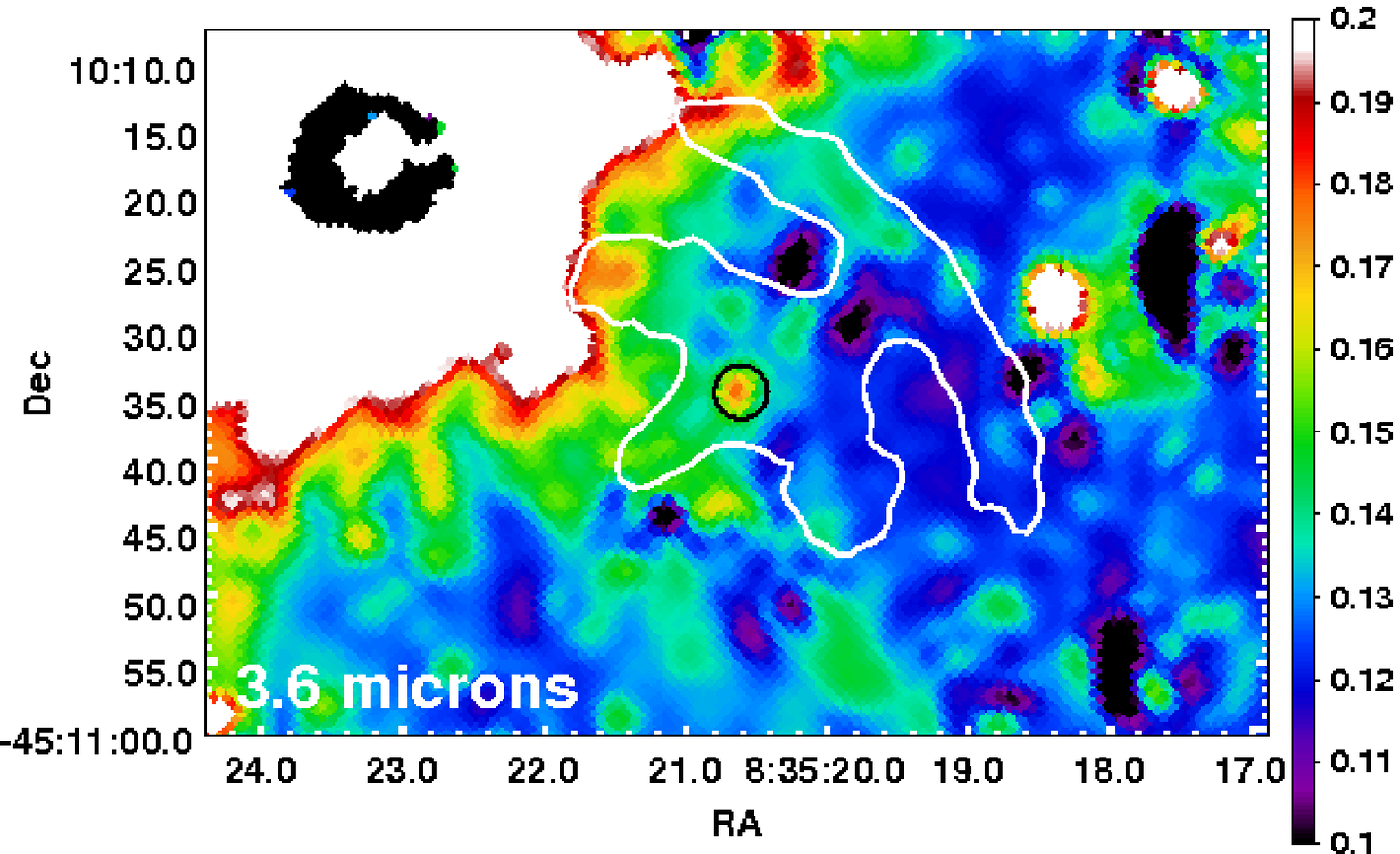}} 
\put (-1.51,0) {\includegraphics[scale=0.52, clip]{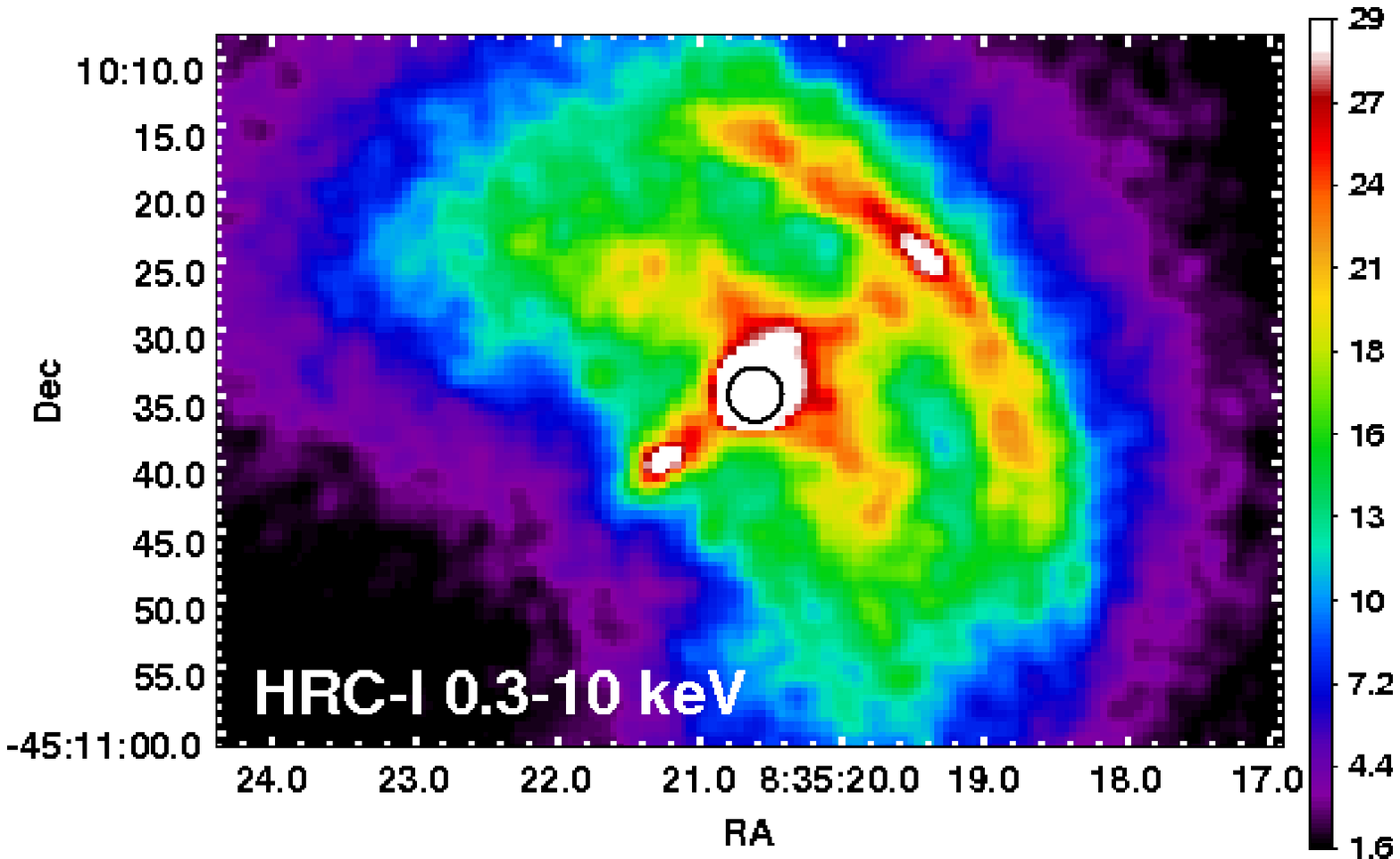}} 
 \end{picture}}
 \caption{ The  $\sim$ 50\arcsec$\times$85\arcsec~fragments of the star-subtracted  3.6  $\mu$m  
 ({\sl top})  and the X-ray  ({\sl bottom})  images obtained with the \textit{Spitzer}/IRAC (AOR 11542784) and \emph{Chandra}/HRC-I, 
 containing the Vela pulsar and  the brightest  part of its  torus-like PWN. The circle
  marks the pulsar position.  
 The  contour indicating the  position of the two bright arcs and south-east jet
 of the X-ray PWN at the bottom panel is overlaid  on the mid-IR image.     
 }
 \label{fig:vela-ima-2}
 \end{figure}

Summarizing this part,  
the optical/near-IR  counterpart 
of the Vela pulsar is firmly established based on its peculiar spectrum, proper motion, 
and pulsations with the pulsar 
period. Therefore, the 1$\sigma$ agreement of the mid-IR counterpart candidate position 
with the near-IR counterpart coordinates,  obtained by us using both    
the relative and absolute astrometries\footnote{We got the same results 
considering another  available optical images   of different epochs  as  reference frames.}, 
allows us to ignore  the  radio-optical standard systematics discussed above,  
and to conclude that the   source detected in the two IRAC bands   is   likely to be associated with   
the pulsar. To confirm the identification by the mid-IR source  
proper motion measurements, higher spatial resolution and/or new \textit{Spitzer}  mid-IR observations 
at later epochs are necessary.
\begin{figure*}
 \setlength{\unitlength}{1mm}
\resizebox{11cm}{!}{
\begin{picture}(150,70)(1,0)
\put (-46,0) {\includegraphics[scale=0.59, clip]{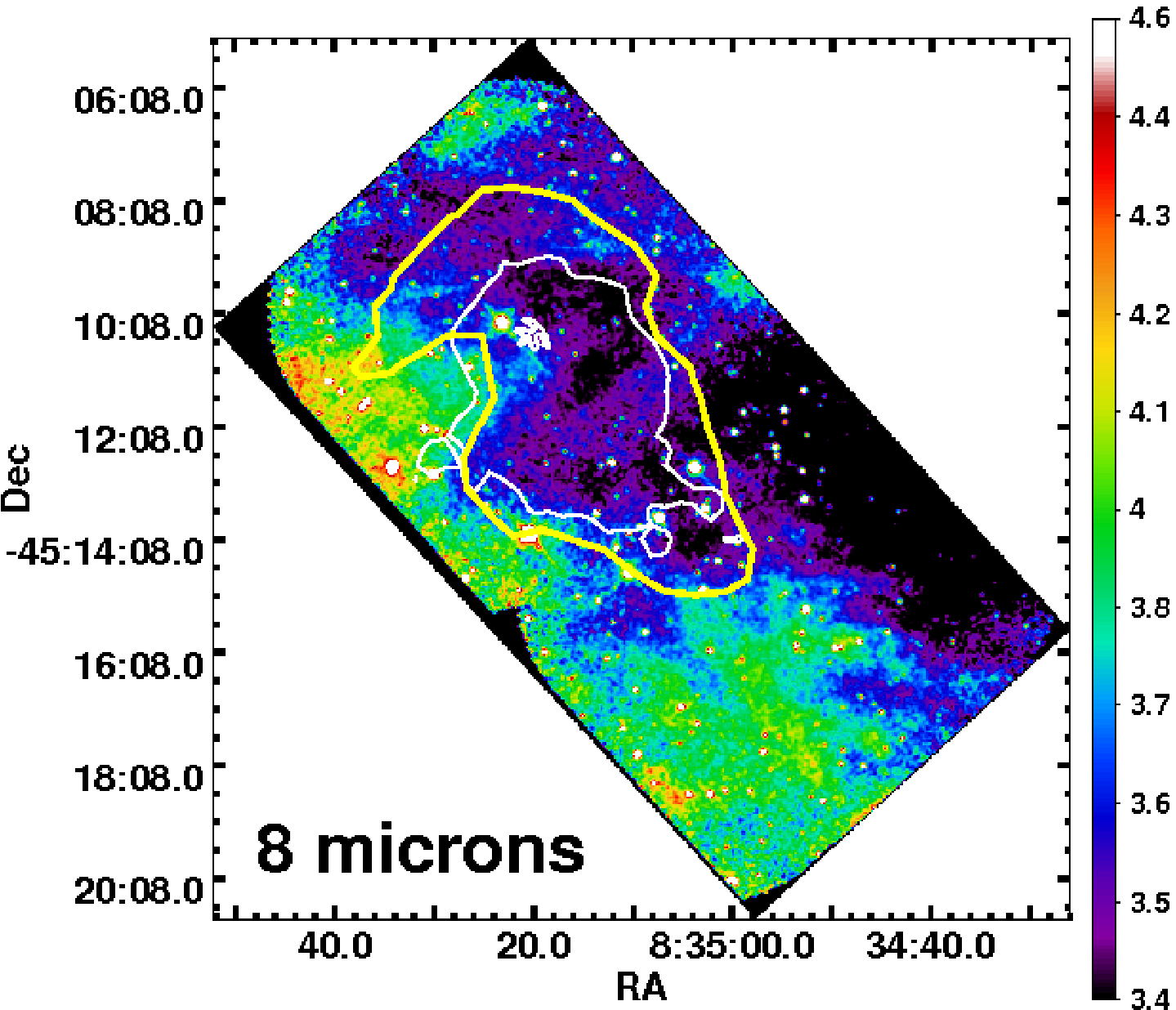}} 
\put (46,0) {\includegraphics[scale=0.53, clip]{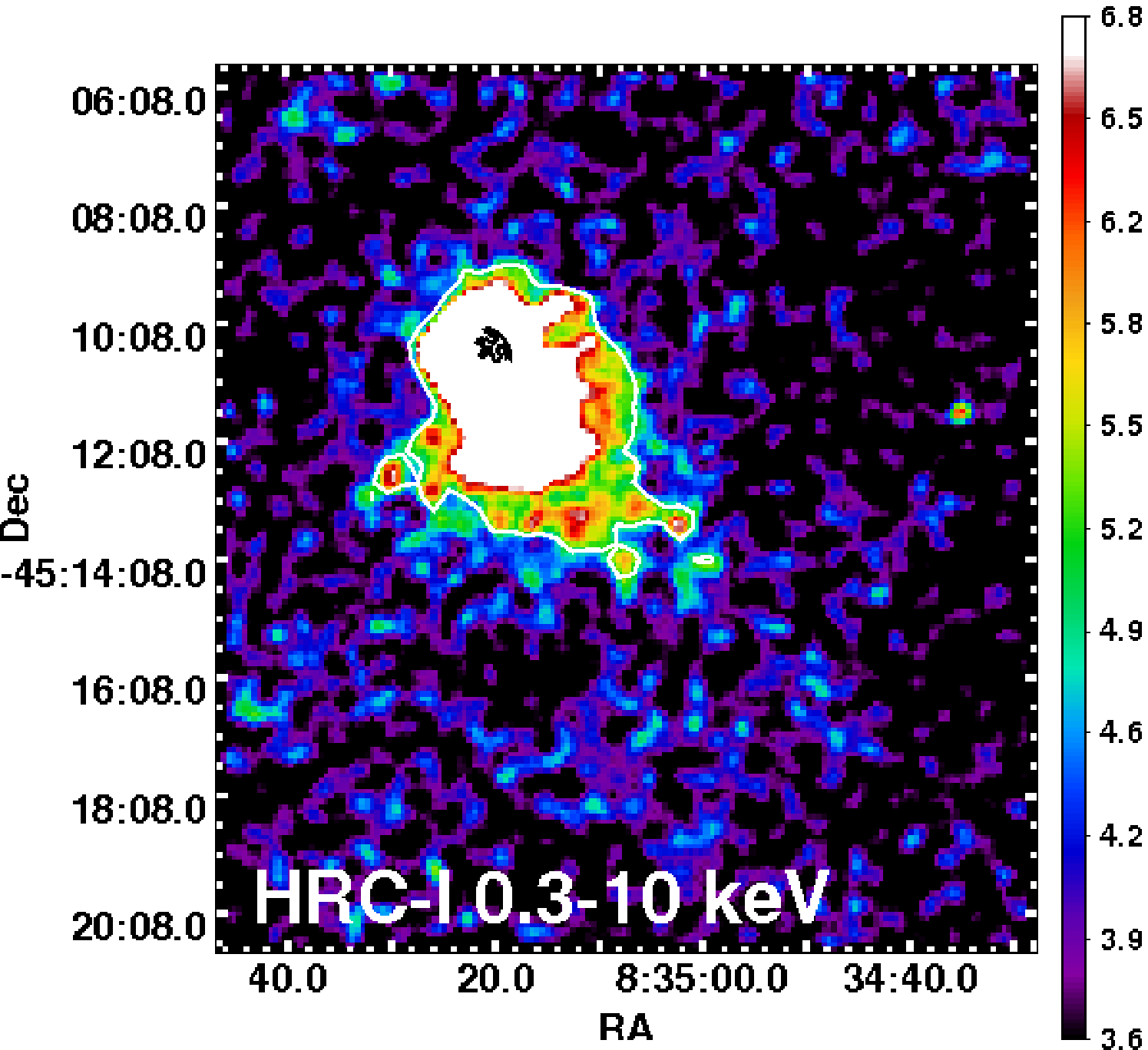}} 
 \put (130,0) {\includegraphics[scale=0.45, clip]{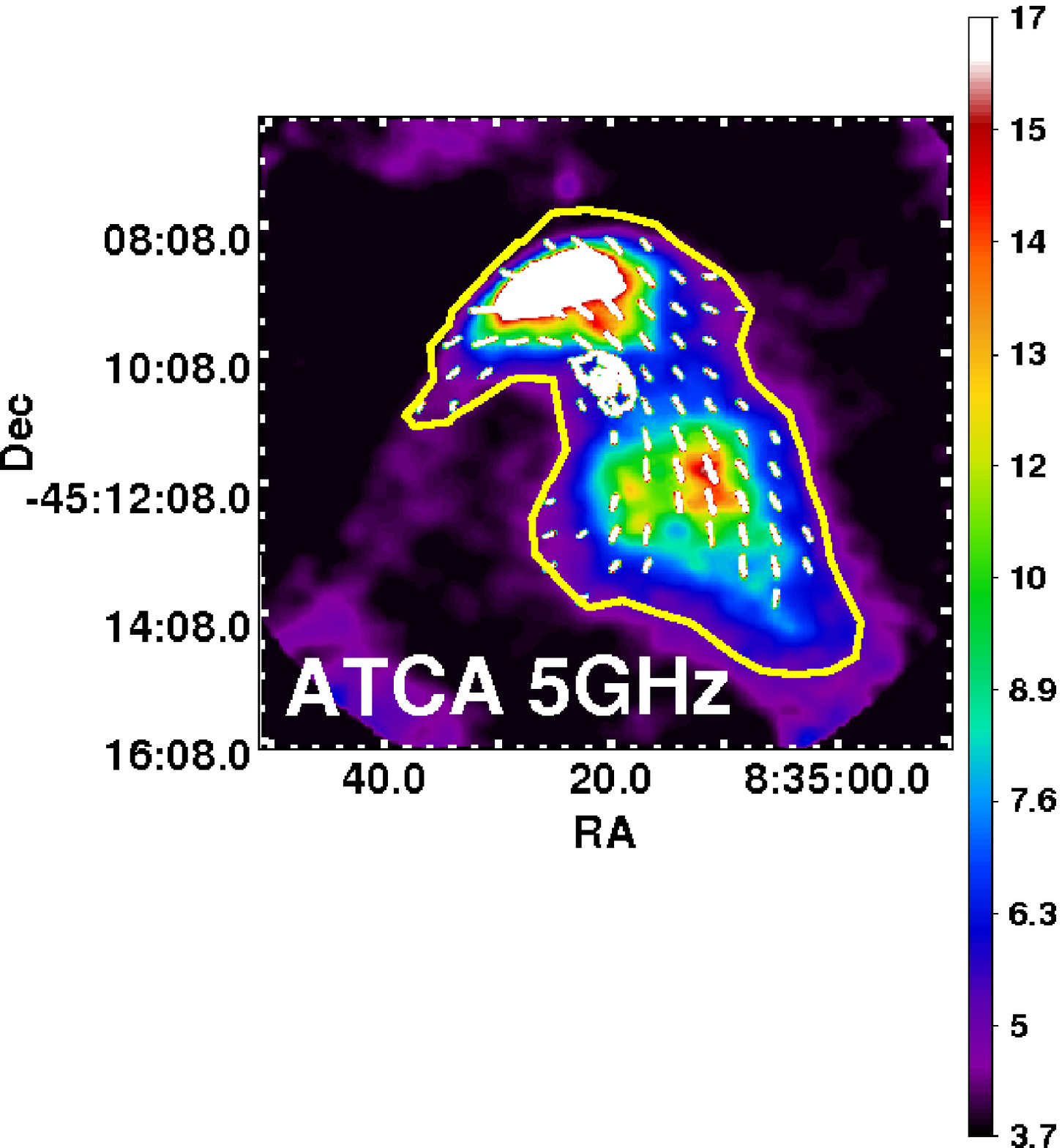}} 
 \end{picture}
 }
 \caption{ The large-scale structure of the Vela plerion at 5 GHz, 8 $\mu$m, and in  soft X-rays, 
as seen with the ATCA ({\sl right} \citep{Dod2003a}), \textit{Spitzer}/IRAC 
 (AOR 11374848) ({\sl left}),  
 and \textit{Chandra}/HRC-I ({\sl middle}).   
 The contours  in the 8 $\mu$m image are from the 5 GHz and X-ray images.  
 The central black and white contours in the X-ray and radio/mid-IR images 
 indicate the location of the internal torus-like  
 structure of the PWN  zoomed  in Fig.~2. Ticks in the 5 GHz image show   magnetic 
 field lines.}
 \label{fig:vela-ima-3}
 \end{figure*}         

  \subsubsection{Searching for  the Vela PWN}
  {\sl Immediate pulsar vicinity.} 

\noindent Previous attempts to find any optical counterpart of the Vela PWN 
have been unsuccessful \citep{mign03}.   
In the  near-IR \textit{H} band  image of Fig.~\ref{fig:vela-ima-1} one can resolve   
a nearby knot-like source 'o1' north-west of the pulsar and a faint  structure, 'counter-jet', 
extended by a few arcsec just south-east of the pulsar \citep{shib03}.  
Both are projected on the pulsar  X-ray jet/counter-jet structures,     and may be associated with the PWN. 
We do  not detect them in the mid-IR. A possible reason is that the second structure 
can be hidden in the pulsar-candidate source wings due to a lower spatial resolution 
of the \textit{Spitzer}/IRAC  as compared to the VLT/ISAAC (see Sect. 3.1.1).  
The  more distant knot could be resolved, however it is not detected. 
It is also not visible in any available optical images obtained at different
epochs. If it is not an artefact, it may be a time variable source.
  
To search for  PWN structures at larger scales,  we  performed PSF subtraction  
of background stars  from the \textit{Spitzer} images making use of  the  IRAF {\sl daophot/allstar} tool.   
The results for the brightest  inner torus-like part of the  PWN are shown in  Fig.~\ref{fig:vela-ima-2},  
where we compare the star-subtracted 3.6  $\mu$m image with the archival \textit{Chandra}/HRC-I 
X-ray image\footnote{Obs. ID 1996, date 2002-01, exposure  49.5 ks, PI D. Helfand.}.  The pulsar counterpart 
candidate was not subtracted and its position is compatible with the X-ray  position of the pulsar marked by the circle.  
Some remains of a poorly subtracted saturated field stars are visible, but  no counterparts 
of the X-ray torus-like  structure with jets and arcs are detected.   We have not found them  in other 
\textit{Spitzer} bands as well.  
 
 \bigskip
 \noindent{\sl Plerion.}     
\newline The inner torus-like PWN, shown in Fig.~\ref{fig:vela-ima-2}, is only a tiny part 
of the Vela plerion,  
that  extends  in the radio and X-rays  to several arcminutes. The respective field is shown  
in Fig.~\ref{fig:vela-ima-3}, where we compare  the  8 $\mu$m,  
soft X-ray, and 5 GHz images. We reproduced the 5 GHz image  from fig.~10 of \citet{Dod2003a}.  
To reveal the plerion X-ray structure the \textit{Chandra}/HRC-I 
image is binned by 16 pixels and smoothed with 3 pixel Gaussian kernel.
The image intensity scale is also different, as compared 
to that of Fig.~\ref{fig:vela-ima-2}, to reveal fainter emission from outer PWN regions,  
while its inner torus-like part is indicated by the contour taken from Fig.~\ref{fig:vela-ima-2} to demonstrate 
its small extent as compared to the whole plerion size.
We have applied contours from the 5 GHz and X-ray images on the 8 $\mu$m image.   
Contour taken from the X-ray image  
shows the outer boundary of the X-ray plerion, where it merges with backgrounds, as it is seen with the HRC. 
Contour taken from the 5 GHz image indicate the outer boundary of the radio plerion, which consists  
of two bright,  the north and south, radio 'lobes',  located symmetrically with 
respect to the inner PWN symmetry axis. 
The inner PWN  X-ray contours are 
overlaid on the 5 GHz image, though this part of the PWN is  not detected in the radio range.    
In contrast to the  torus-like structure, whose symmetry axis is directed north-west 
approximately along the proper motion of the pulsar,  the plerion   is mainly extended 
south-west.  
 
As well as in the radio image, the pulsar and its inner PWN are not detected at 8 $\mu$m.  
However, the outermost east boundary of the large scale 
plerion and its  south-west extension in the radio and X-rays likely correlate 
with the west boundary 
of a bright extended emission structure in the mid-IR, which  also  extends south-west and,  
possible, originates from the dust formed 
from the remnant ejecta  and heated by the PWN. 
 In other SNRs at longer wavelengths  
the radiation from such  extended structures  is typically dominated  
by  continuum emission from the dust, that can significantly contribute  
to the radiation of a remnant  at 8 $\mu$m as well \citep[see, e.g.,][]{rho09}. 
 Indeed, the similar but brighter structure is visible at  24   and 70  $\mu$m images 
 obtained with the MIPS. 
 However, they  have much smaller and not representative FOVs, and we do not show them here.
 Nevertheless, the fact, that the extended mid-IR emission is visible in the images obtained 
 with different \textit{Spitzer} detectors in three different bands, at different AORs 
 and observation conditions, suggests   
 that it is real, but not a result of an artificial background variation.
  At shorter wavelengths the extended emission disappears, as expected for 
  a rather cold dust.   
 
The centre of the associated  with the plerion TeV source  HESS J0855$-$455  \citep{ahar2006} 
is located tens arcminutes south-west and is outside the frames shown in Fig.~\ref{fig:vela-ima-3}. 
It is mainly extended in the same south-west direction as the X-ray/radio plerion. 
Unfortunately, there are no \textit{Spitzer} images to combine a complete large scale picture 
of the extended mid-IR emission,  around  the Vela plerion including the  TeV source. 
Nevertheless, a remarkable coincidence of its complicated east    
X-ray and radio boundary with the west boundary of the extended mid-IR emission 
suggests, that we likely see  an interface between the pulsar wind and the  SNR ejecta  
supported by a pressure balance between these two different fractions of the remnant. 
It is most clearly visible in the north-east part of the plerion, where a bright 
mid-IR protrusion fills a deep  nebula bay between its north and south 
radio lobes.     
 As seen from the radio image of Fig.~\ref{fig:vela-ima-3}, the magnetic field lines also lie  
 along  the protrusion/bay boundary, possible contributing to the pressure balance 
 in this region. 
 We see no extended mid-IR emission along the west plerion boundary. However, the magnetic field  
lines in the 5 GHz image are concentrated towards this boundary and mainly directed along it. 
 The magnetic fields are stronger there and may play a key role in the shape 
 and stability of the west part of the plerion. 
\begin{figure*}
 \setlength{\unitlength}{1mm}
\resizebox{11.9cm}{!}{
\begin{picture}(120,122)(1,0)
\put (-28,54) {\includegraphics[scale=0.53, clip]{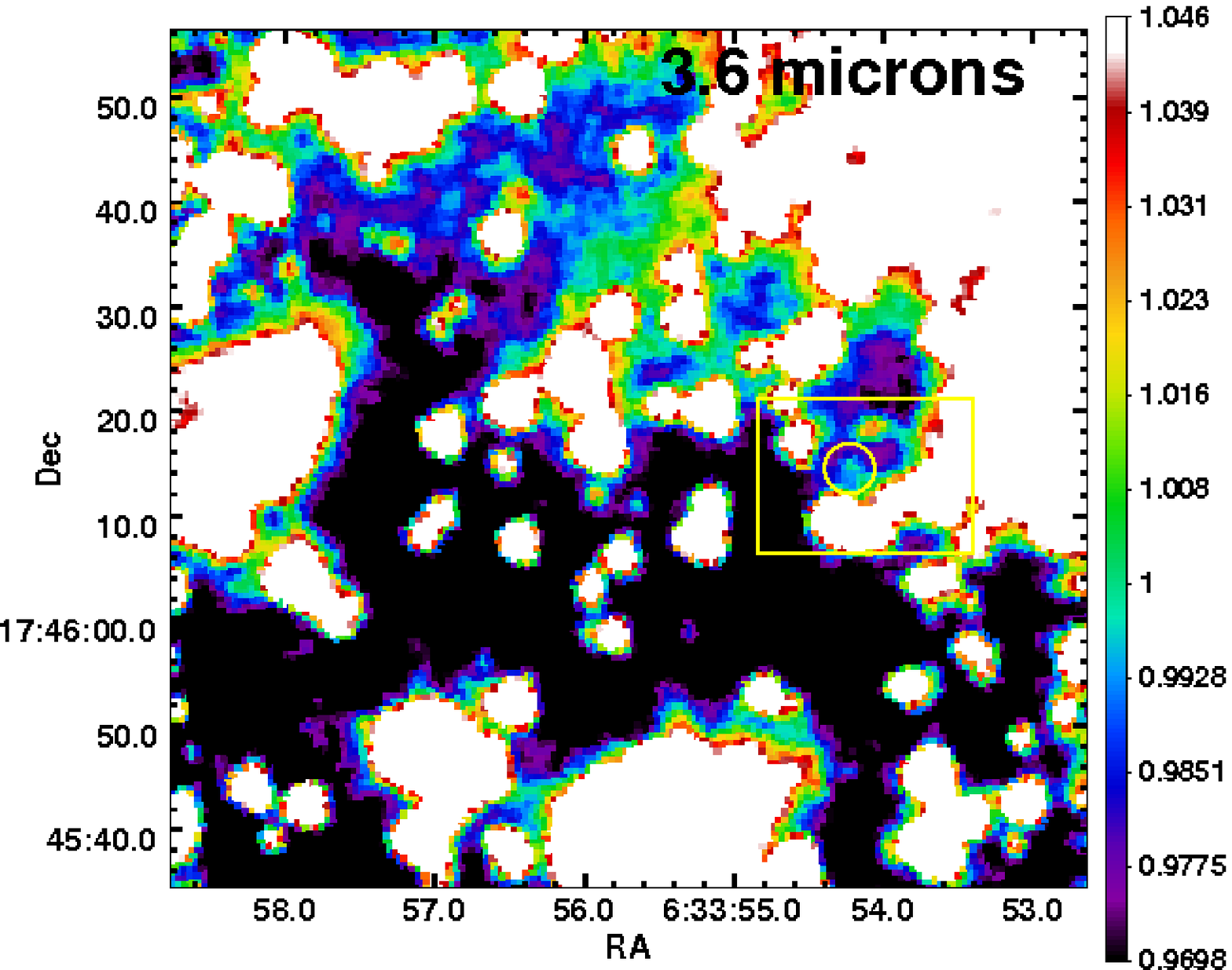}} 
\put (62,54) {\includegraphics[scale=0.53, clip]{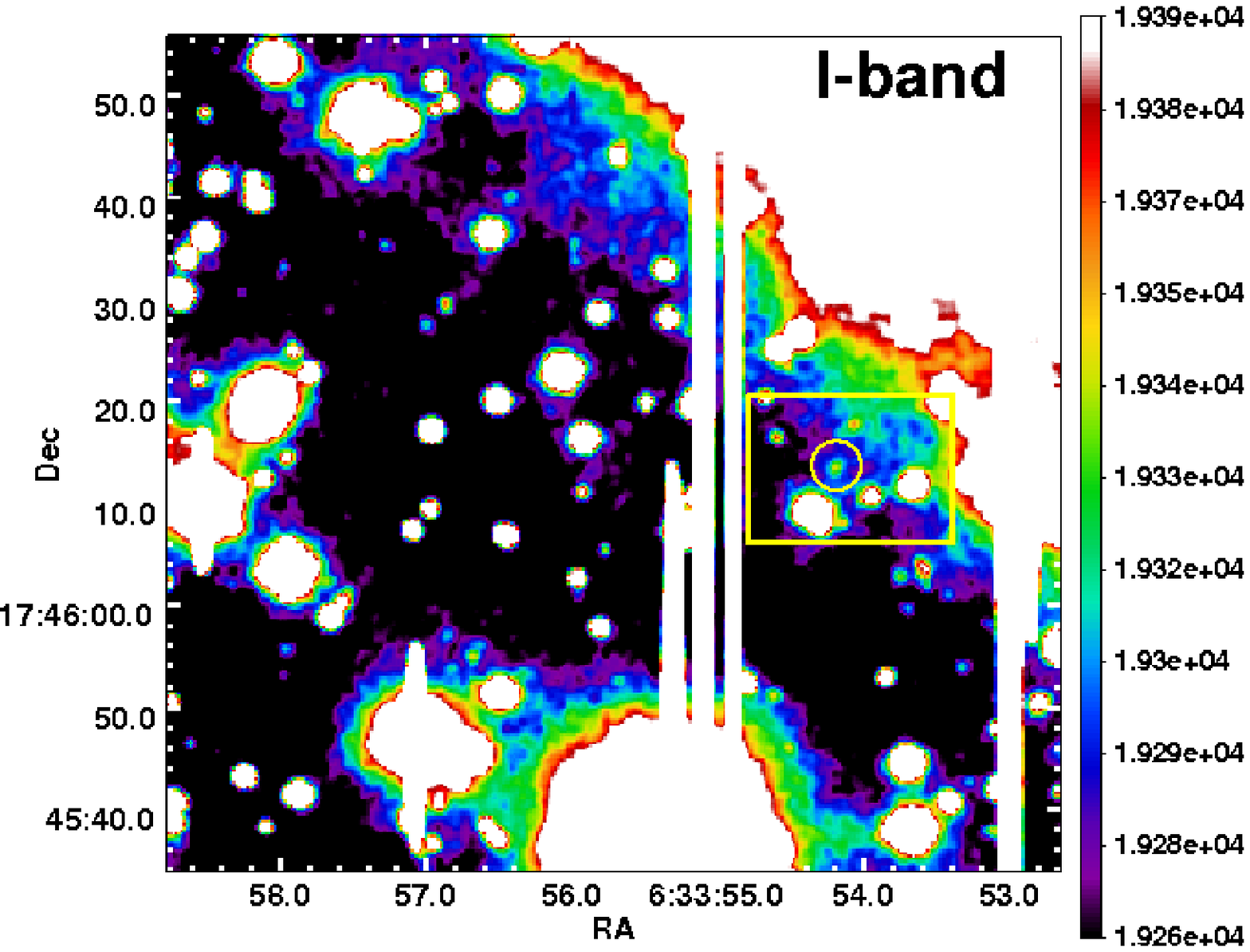}} 
\put (-29,0) {\includegraphics[scale=0.5, clip]{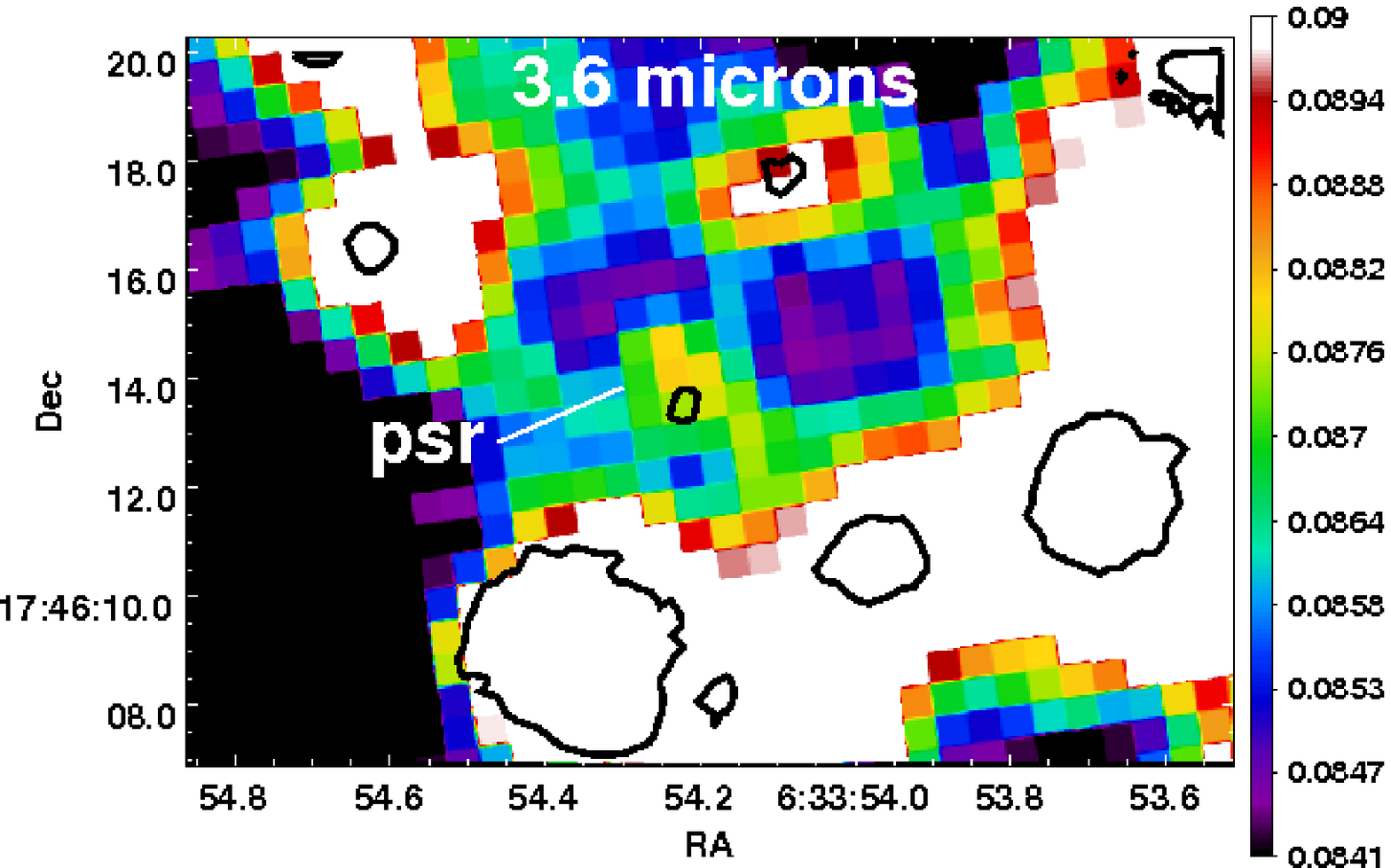}} 
\put (62,0) {\includegraphics[scale=0.5, clip]{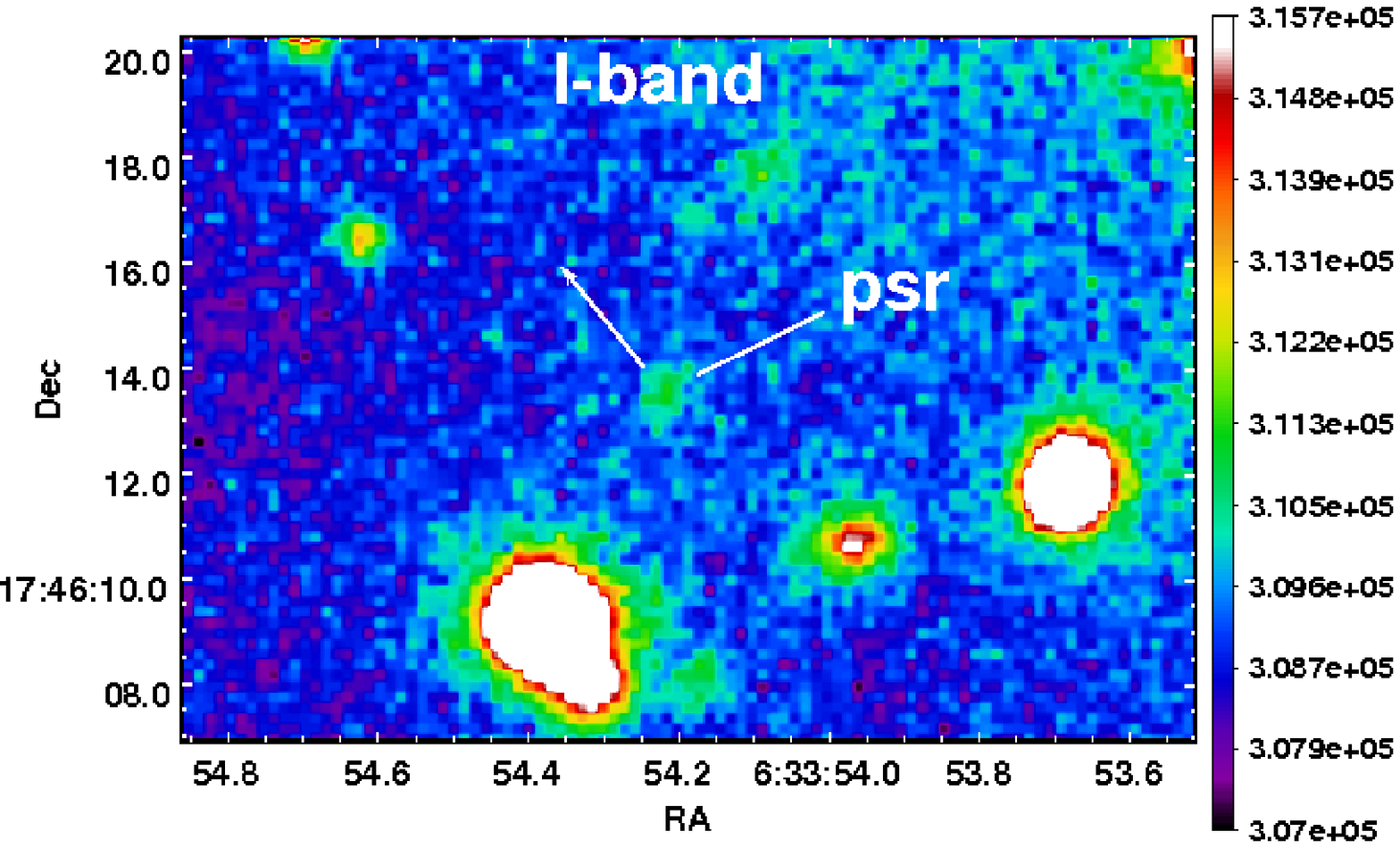}} 
 \end{picture}
 }
 \caption{ {\sl Top panels:} 
 An $\sim$ 1\farcm2$\times$1\farcm2~overview of the Geminga field, as seen  with 
 the \textit{Spitzer} (AORs 19037696+19037952)  
  at 3.6 $\mu$m  ({\sl left})
   and   Subaru  in the \textit{I} band ({\sl right}).  The \textit{I} band image is smoothed 
   with four pixel Gaussian kernel, that is about the mid-IR pixel size. 
   It is partially affected by  several vertical  stray light artefacts 
   form oversaturated stars.   
   Circles mark the Geminga 
   optical counterpart and its mid-IR counterpart candidate. 
   The  $\sim$ 15\arcsec$\times$20\arcsec~box region of the pulsar neighbourhood 
   is enlarged in  
   {\sl Bottom panels}, where  
 the Geminga  counterpart  is indicated by  a line, and the arrow shows 
 the direction of its proper motion. 
 The contours in the  bottom 3.6 $\mu$m image, where the only AOR 19037952 
 is implemented,  are from the \textit{I} band image. 
 }
 \label{fig:gem-ima-1}
 \end{figure*}

  \subsection{Morphology of the Geminga field} 
  \subsubsection{Possible identification of the Geminga pulsar}   
In Fig.~\ref{fig:gem-ima-1} we compare fragments    
of the Geminga  field images obtained in the optical \textit{I} band  on  January 2001 
with the Subaru telescope and  
at 3.6 $\mu$m with the  \textit{Spitzer}/IRAC.    
The  \textit{I} band image is taken from \citet{shib06}.
The  pulsar optical counterpart is indicated by the yellow circle 
in the top panels   
 and by the thin line in the bottom ones, where an overview of the field and 
 pulsar immediate vicinities  are shown, respectively.    
 The direction of the pulsar proper motion  
of 178.2$\pm$1.8 mas yr$^{-1}$  with a position angle of 52\fdg9$\pm$0\fdg4 
\citep{fah07}  is shown by the arrow in the bottom \textit{I} band image.  
For the 3.6  $\mu$m observation  epoch the expected  displacement of the pulsar 
from its \textit{I} band position  
is $\approx$ 1\farcs22. 
It is consistent at 1$\sigma$ astrometric uncertainty level (0\farcs4) 
with the positions of a faint mid-IR source  marginally detected 
 at 3.6 $\mu$m, that is also marked as 'psr'.  
 The position of the source itself is defined with a twice worse accuracy 
 of $\approx$ 0\farcs8, as compared to the Vela counterpart candidate case. 
 The main reason is its faintness  and a poor approximation of 
 the object brightness distribution by   a point-like source profile. 
 Nevertheless, it  is fully compatible with  the pulsar proper motion path, 
 and  can be considered  as a likely mid-IR counterpart of Geminga.  

Despite a worse spatial resolution of the \textit{Spitzer} compared to the Subaru, 
as in case of the Vela pulsar, practically all background optical objects in 
the frames  shown in Fig.~\ref{fig:gem-ima-1}  can be identified in the mid-IR 
and vice versa. This is underlined  by  black contours from the \textit{I} band image,
which are overlaid on the bottom mid-IR image. We have not  found also any additional 
background object in immediate vicinity of  Geminga in  the near-IR image of the field obtained 
 with the \textit{HST} in the adjacent F160W band \citep{shib06}. Considering much 
 larger FOV, as in the top panels 
of Fig.~\ref{fig:gem-ima-1}, we found only a few new faint 
mid-IR sources, which are not visible in the \textit{I} band and located  far away of the pulsar.   
  Hence, the appearance of a new background source at the expected Geminga position 
  in the mid-IR is very unlikely. This supports the correct identification of the pulsar.  
\begin{figure*}
 \setlength{\unitlength}{1mm}
\resizebox{10.cm}{!}{
\begin{picture}(120,79)(1,0)
\put (-45,0) {\includegraphics[scale=0.65, clip]{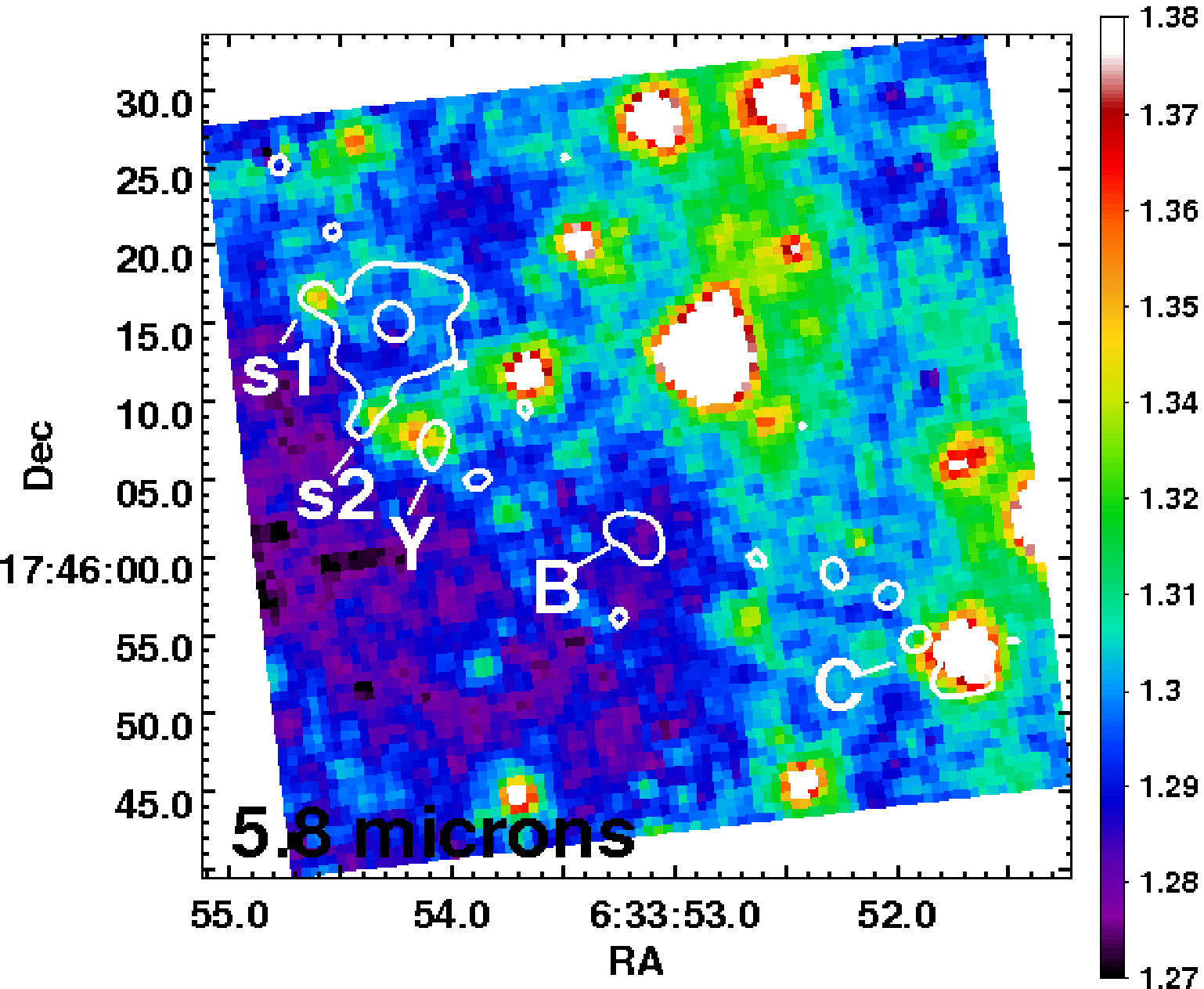}} 
\put (63,0) {\includegraphics[scale=0.65, clip]{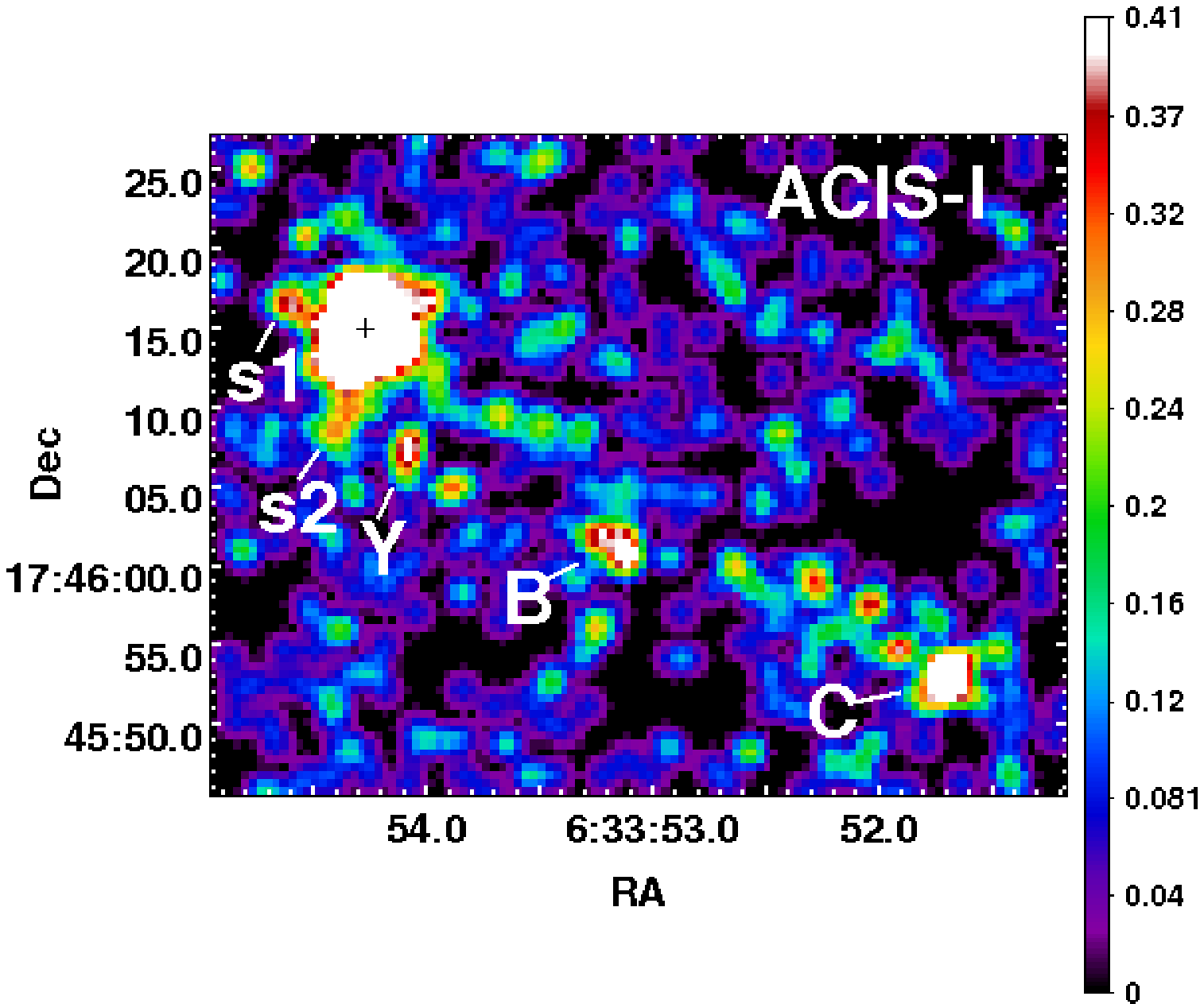}} 
 \end{picture}}
 \caption{The large scale overview of the Geminga field at 5.8  $\mu$m ({\sl left}) 
 and in X-rays ({\sl right}). 
 The contours in mid-infrared image are from the X-ray image. Bright components of 
 the  clumpy Geminga tail, as well as 
 the two background IR/optical sources, 's1' and 's2', are marked.  Geminga, marked by the cross 
 in the X-ray image, 
 is not detected in this infrared band. A bright X-ray blob B and other 
 less bright blobs  have   no infrared counterparts. While the blob C at the end 
 of the X-ray tail  is likely associated 
 with a bright star,  and  the  blob Y may be associated with a red background source 
 not visible at shorter IR/optical wavelengths.}
 \label{fig:gem-ima-4}
 \end{figure*}
  
Unfortunately, the suggested counterpart candidate, 
is detected only at $\sim$ 2$\sigma$ level  
and only in the one \textit{Spitzer} band. 
Combining  the data from the two AORs (19037696+19037952),  
as is in the top panel of Fig.~\ref{fig:gem-ima-1},   
has not allowed us to increase the significance of its  
detection due to a  high Zodiacal background  
in the first AOR, though the source is seen at the same coordinates 
in both AORs (see Sect. 2.1). Therefore, this source has to be considered 
with a caution and  needs a confirmation by deeper observations. 
Nevertheless, summarizing all pros and contras, we will assume below,  
that the source is the real  counterpart of Geminga, and  estimate how this may affect     
the broadband spectrum of the pulsar.        

 \subsubsection{Searching for the Geminga PWN}  
Deep X-ray observations have allowed to find fragments of a bow shock 
and an axial tail  
of Geminga at the distances   of several arcminutes behind the pulsar 
\citep{caraveo2003, pavlov2010}. 
We have not found any significant 
mid-IR counterpart of the bow shock  at these  scales. 
At the same time, the tail was found to consist of a time-variable clumps appearing and disappearing in various 
places from time to time \citep{pavlov2010}, 
and a few possible counterparts of  the  clumps  can be considered.         
An overview of the pulsar field containing the whole extent of the axial tail  
is shown in    Fig.~\ref{fig:gem-ima-4}, where we compare the mid-IR 5.8 
$\mu$m  and  ACIS-I\footnote{Obs. ID 7592, date 2007-08, exposure  72.1 ks, PI G. Pavlov.} X-ray images.  To go as deep as possible, 
the mid-IR image was combined 
from the two longest observations  listed in Table 1 (AORs 19037696 and 19037952). The brightest fragments of 
the X-ray tail detected at the ACIS-I observation epoch are marked by B and C, in \citet{pavlov2010} notations, 
and yet another one by Y. The X-ray contours are overlaid on the mid-IR image. 
 
There is no any significant mid-IR counterpart for the X-ray time-variable B substructure. 
It was not seen in the early ACIS-S\footnote{Obs. ID 4674, date 2004-02, exposure  18.8 ks, PI D. Sanwal.} 
X-ray image and, hence, has been born at a time 
within a  3.5 yr interval between  the two  X-ray observations. 
If we assume that appearing a bright X-ray clamp is accompanied 
with an optical/IR emission, then  
the absence of B in the mid-IR suggests, that it was 
created either after the \textit{Spitzer} observations and its age 
is $\la$ 9 months, 
or
it has flashed before these observations, but the   lifetime   
of  the particles responsible for possible  mid-IR afterglow associated with the X-ray flash  
is much shorter than that for the particles  radiating in X-rays.

The blob C  nicely coincides with a background  stellar source, as has been 
noticed by \citet{pavlov2010}.  
We can add, that in the mid-IR  this source is  not a single  but  likely 
a bland of one bright an 2--3 fainter stars. 
 
The blob Y positionally coincides with a faint background source in the 5.8 $\mu$m image. The
source is not visible at shorter wavelengths and hardly visible at 8 $\mu$m.
This blob is close to a bright blob A, that was clearly visible  in the  
early ACIS-S image \citep[see,][]{pavlov2010}   and  not seen in the ACIS-I image shown here, as well as  in any available IR/optical image. 
It is not clear, whether this is a remnant of A, that has faded and moved by 
$\sim$ 3\arcsec~to the east from its initial 
X-ray position, or it is an independent substructure, that was flashed later 
in the time interval between the two X-ray observations. 
Additional X-ray and mid-IR observations 
are necessary to understand whether  the Y blob and the suggested IR counterpart 
are variable and have the same origin or not.  If we indeed detected  this blob almost simultaneously in the X-ray and mid-IR, 
it could be younger than the blob B. 
The rest fainter and less extended X-ray sources possible related to the extended 
tail are not identified in the mid-IR.

Considering immediate vicinity of the pulsar, \citet{pavlov2010} discussed  
a bow shock structure of the X-ray PWN with  possible 'forward-jet', which is  directed 
north-east of Geminga along its proper motion. 
As seen from Fig.~\ref{fig:gem-ima-4}, the forward-jet substructure, marked as 's1', 
is likely  not  related to the PWN,  but is the X-ray counterpart of a red 
star-like background  optical object, that is also detected in the near-IR and  optical 
images (cf. Fig.~\ref{fig:gem-ima-1}). In the deep ACIS-I image its profile overlaps 
with the outermost region of the nebula producing a false impression on 
the presence of a jet. The object s1 is also visible as  a point-like isolated 
object in the ACIS-S image,  where the exposure was short.      
A similar situation is for an apparent  south-east 'protrusion' seen  
only in the ACIS-I image and marked as 's2'.  
It is likely to be a background source as well, that is brighter in the optical/IR, 
but fainter in X-rays as compared to 's1'.                 
\begin{table}
\caption{The results of the aperture photometry 
for the presumed mid-IR counterparts of the Vela and Geminga pulsars. The 3rd column shows the magnitudes and upper 
limits measured using optimal aperture  radii presented in the 2nd column. The presented aperture correction factors  
are for the flux.    
}
\begin{tabular}{cccccc}
\hline\hline
  $\lambda_{eff}$ & apert.& mag. & apert. &     mag.                      &  log F                               \\
                           &  radi-      &  measur-          & correc-      &  correc-                  &                \\
            &   -us      &    -ed$^a$       &  -tion         &              -ted$^a$               &              \\
           &         &                 &    factor        &                             &              \\
($\mu$m)            &   (pix)      &  (mag)         &           &  (mag)                           & ($\mu$Jy)             \\
\hline
                         \multicolumn{6}{c}{Vela}                                             \\
 \hline
 160         &  8 &  $\ga$ 2.34&    1.9                & $\ga$ 1.64                          &$\la$ 4.5                      \\
 70         &  8.7 &  $\ga$ 6.3&    1.2                & $\ga$ 6.1                          &$\la$ 3.44                      \\
 24        &5.3& $\ga$ 11.41&      1.16                    &  $\ga$ 11.25                      &$\la$ 2.35                   \\ 
 8.0         &10 & $\ga$ 15.65      &    1.07     &   $\ga$ 15.58                   &  $\la$ 1.70                        \\
 5.8        &   4  & 16.73(27)   &    1.38                   & 16.38(27)                            & 1.51(11)           \\
 4.5        & 10  &  $\ga$ 16.90  &  1.06       &    $\ga$ 16.84                          &$\la$ 1.52                                  \\
 3.6       &    4  &  18.69(22)   &    1.21                    & 18.48(22)                            & 1.06(09)            \\
 \hline 
                                         \multicolumn{6}{c}{Geminga}                          \\ 
\hline
24         &5.3 & $\ga$ 11.25      &    1.16     &   $\ga$ 11.09                   &  $\la$ 2.42   \\
8.0         &10 & $\ga$ 16.3      &    1.07     &   $\ga$ 16.23                   &  $\la$ 1.44                        \\
 5.8         &10 & $\ga$ 18.9      &    1.06     &   $\ga$ 18.84                   &  $\la$ 0.52                        \\
 4.5        & 10  & $\ga$ 20.48  &  1.06       &   $\ga$ 20.42                           &$\la$ 0.09                                  \\
 3.6      &  4   &      21.77(78)   &  1.21           &    21.56(78)                 &   -0.18(31)                              \\
\hline
\end{tabular}
\label{t:phot}
\begin{tabular}{ll}
$^a$~numbers in brackets are 1$\sigma$ uncertainties referring & \\
to last significant digits quoted & \\
\end{tabular}
\end{table}
 \subsection{Photometry of counterpart candidates}
To minimize the flux contamination of the detected  pulsar counterpart candidates from nearby background objects  
the aperture photometry was performed on the star-subtracted IRAC images.

For  magnitude estimates we used IRAF {\sl daophot} tasks 
with zero points given in the IRAC  Instrument Handbook\footnote{see, 
http://ssc.spitzer.caltech.edu/irac/iracinstrumenthandbook/}. Prior  photometric measurements, 
we converted each image from MJy steradian$^{-1}$ to DN units using the conversion factors 
from image fits headers. In this case the {\sl daophot } magnitude error output can be directly used 
for the flux uncertainty estimates.  To find an optimal aperture radius based on the maximal signal to noise ratio,   
photometric growth curves were obtained and used. 
For  the pulsar  candidates at 3.6  and 5.8 $\mu$m the optimal circular aperture radius 
was 4 pixels (at 0\farcs6 pixel scale) with the background annulus and dannulus 
in the ranges of 5--6 and 4--6 pixels, respectively. 
The measured fluxes were multiplied by  aperture correction factors taken from the IRAC Data 
Handbook\footnote{IRAC Data Handbook, table 5.7.}. 
We have performed  the PSF photometry as well and got 
the results consistent  
with the aperture photometry at 1$\sigma$ level. 

Conservative 3$\sigma$  point source magnitude upper limits can be placed  for  the counterpart candidates in 
the mid-IR bands where the pulsar fields have been observed but no  candidates  were  found.  To do that, the circular 
aperture radii were selected encapsulating $\ga$ 80 \% of a potential point source flux, the apertures were centred 
at the pulsar positions, and  standard deviations and inherent flux variations within these apertures were used.  
The  magnitudes were converted into fluxes in physical units and the results are summarized in Table \ref{t:phot}.
For Geminga at 3.6 $\mu$m, where the counterpart candidate is visible only at 2$\sigma$ level,  the data 
from the AORs 19037952 and 19037696 (Table \ref{t:obs})  provide consistent results. However, the flux uncertainty 
is smaller for the former AOR, where the  Zodiacal  background is lower, and we use this result for our further analysis.

\subsection{Multiwavelength spectra}
\begin{figure}
\begin{center} 
\includegraphics[width=8cm, clip]{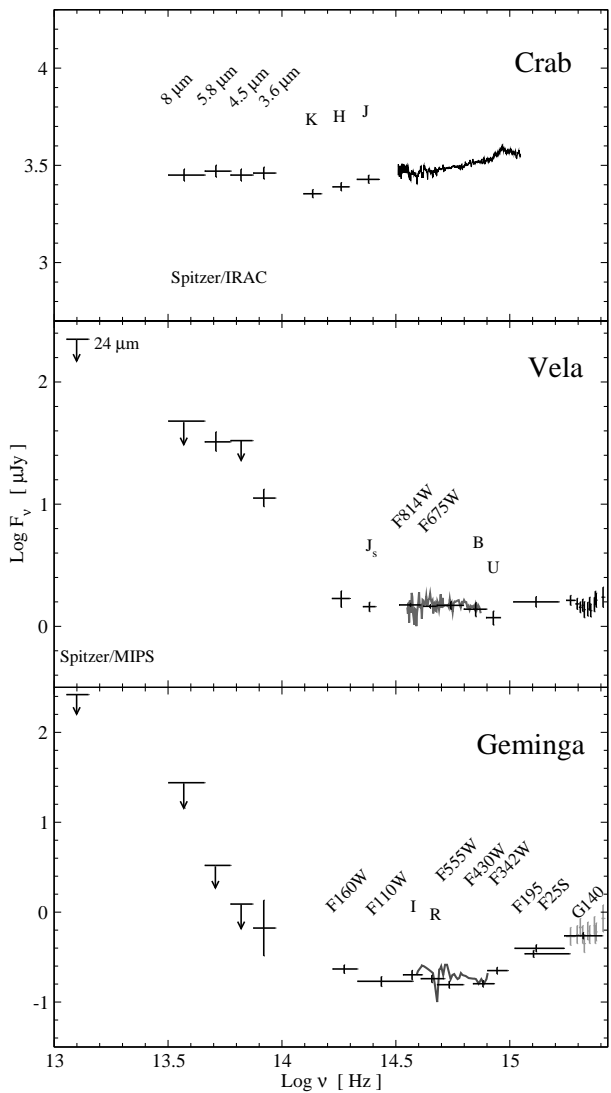}
 \end{center}
 \caption { Unabsorbed spectra of the Crab, Vela, and  Geminga pulsars  from  
 the mid-IR to the optical--UV. 
 The Crab data are from \citet{sol09}. The Vela  UV, optical, and near-IR 
 data are from \citet{Romani2005},  \citet{mign07},  
 and \citet{shib03}, respectively. The Geminga UV and optical/near-IR  
 data are from \citet{Kar2005} 
 and \citet{shib06}, respectively, the optical spectrum is from \citet{Martin1998}.}
 \label{fig:mw}
 \end{figure}

Let us assume 
that our mid-IR identifications are real, and consider,  
how this can affect the spectra of the pulsars.      
Using the measured mid-IR fluxes,   available optical and UV data, 
in Fig.~\ref{fig:mw} we compile  
dereddened multiwavelength spectra 
of the Vela and Geminga pulsars
in the mid-IR-UV range, and compare them with the respective spectrum 
of the Crab pulsar taken from \cite{sol09}. The  interstellar extinctions 
\textit{E(B-V)} are 0.055 and 0.023 mag towards Vela and Geminga, 
respectively \citep{shib03,shib06}. These are small enough, and dereddening corrections 
in the mid-IR bands are negligible as compared to the measured flux uncertainties. 
For instance, the magnitude correction A$_{\lambda}$ at $\lambda$ = 3.6 $\mu$m  is $\la$ 0.01 mag 
for both objects. For clearness,  in Fig.~\ref{fig:mw} we notify the  broad 
bands, where pulsar fluxes were measured, and indicate which ones were obtained  
with the \textit{Spitzer}/IRAC or with the MIPS.     
 
In contrast to the Crab, 
that shows almost a flat spectrum 
in the considered range, the compiled spectrum of the Vela pulsar demonstrates a  strong, 
an order of  magnitude, mid-IR flux excess,  that progressively increases  
with the wavelength, as compared to the optical range. 
Geminga shows a similar, though much less significant, excess, 
as follows from its marginal detection.    
It is important to note, that an intention of the flux increase towards the IR has been noticed  by \cite{shib03,shib06}, 
when both pulsars were identified  in  the near-IR \textit{JH} bands. 
However,  the significance of that was low, and the presence of such a strong  
mid-IR excess  was  difficult to  expect,  based on the flat spectrum of the Crab pulsar, 
that has been so far the only rotation-powered pulsar detected in the mid-IR. 
The mid-IR fluxes of the Vela and Geminga candidates are likely to be in agreement 
with their 
near-IR counterpart fluxes, 
    confirming  the reality of the apparent flux increases towards 
the IR, resulted early  only from the near-IR data.  
 Thus, the compiled  spectra  appear to be real, but not the tentative ones. 
 We consider this 
 as an additional  evidence,  
 that the detected mid-IR emission is indeed associated  with the pulsars.     

Accepting that, we can now update the rotational phase averaged spectral pictures 
for both pulsars in the whole observational  range,  
from the radio to hard $\gamma$-rays 
\citep[cf.][]{shib03,shib06}, including our mid-IR data, and  the data 
obtained recently with the \textit{Fermi} $\gamma$-ray observatory  
\citep{Abdo2010v,Abdo2010g}.     

As seen from Fig.~\ref{fig:mw1}, the flux density of the mid-IR emission from the Vela pulsar 
becomes comparable with the  density of the spectral  bump produced by  thermal emission 
from the atmosphere (atm.) at the surface of the NS visible in soft X-rays. At the same time, within large uncertainties it 
may be compatible with the long wavelength extrapolation of the nonthermal soft X-ray spectral tail detected with 
\textit{Chandra} and the \textit{RXTE} and described by a power-law (hatched region). The mid-IR flux upper limits are still lower 
than flux  densities  of  a coherent  radiation 
at  radio frequencies.

Similar spectral behaviour is observed for Geminga.  The main difference  
is that Geminga is fainter  in  the optical/IR, and the observed mid-IR excess 
is much less significant. 
Nevertheless, the steep IR flux increase towards low frequencies can not be excluded   
by the derived mid-IR flux upper limits. This suggests, that   
at several 10s microns   the mid-IR flux density  might be comparable 
with the soft X-ray thermal blackbody-like emission  bump from the Geminga surface (BB)  
and compatible with the low frequency extension of the nonthermal soft-X-ray spectral tail, 
as in the Vela case.  

However, from the  spectral behaviour of both pulsars shown in Fig.~\ref{fig:mw} and Fig.~\ref{fig:mw1},  
it is most likely that  the \textit{Spitzer} mid-IR observations reveal us a new spectral component in the emission 
of these rotation powered-pulsars, whose origin is different  from  those 
of the rest components   detected early, and, perhaps, from that 
of the Crab in the mid-IR.             
\section{Discussion} \label{discuss}
Subarcsecond positional coincidence of the detected point-like object 
at 3.6 and 5.8 $\mu$m with  the Vela pulsar coordinates   suggests, 
that this object is  likely to be the mid-IR counterpart of the pulsar. 
This is also supported  by the consideration of its spectral energy distribution (SED), 
showing that the measured mid-IR fluxes are compatible with 
the long wavelength spectral excess observed early in the near-IR.   
 The same  is likely to be true for fainter Geminga, with a caution that 
 its  mid-IR counterpart candidate  is  detected only marginally and solely in  
 the  3.6 $\mu$m band. Additional confirmations 
can be obtained from near-IR detections of both pulsars in the \textit{K} band, 
where they have not been observed, and by deeper mid-IR observations, 
to detect the Geminga counterpart with a higher significance and 
to establish both  identifications by  proper motion measurements 
of the counterparts in this range  and checking their  consistence    
with the proper motions of the pulsars.  
Extension of the pulsar observations 
to the far-IR would be also useful in this context.       
 
 \subsection{The nature of the mid-IR excess}       
If our mid-IR identifications  of the two pulsars are real, the question is what is the nature 
of the observed mid-IR flux excesses. 
\begin{figure*}
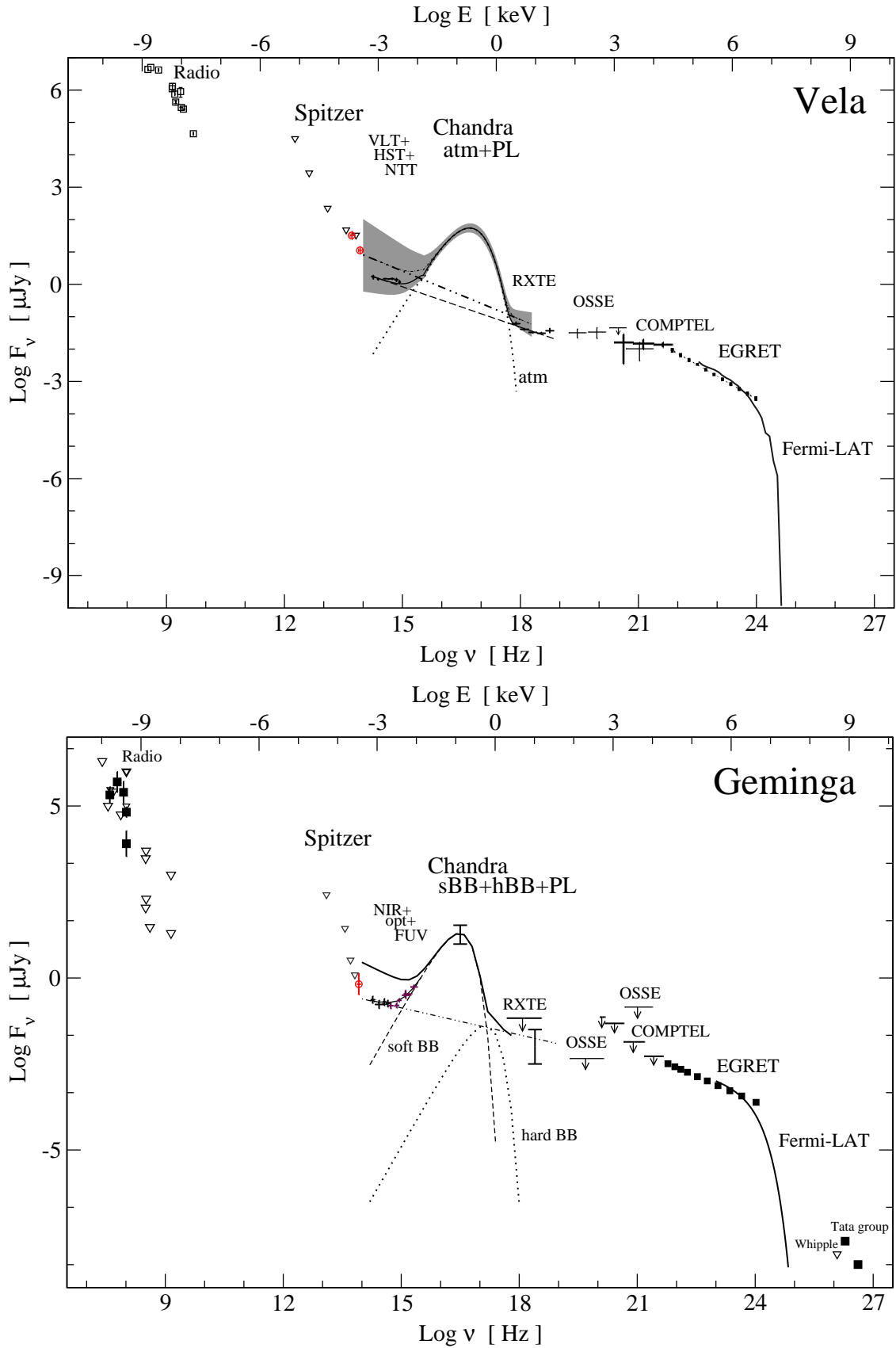

 \setlength{\unitlength}{1mm}
\resizebox{11.5cm}{!}{
\begin{picture}(120,235)(1,0)
\put (-10,120) {\includegraphics[scale=0.58, clip]{fig7.1.col.eps}}
\put (-10,0) {\includegraphics[scale=0.58, clip]{fig7.2.col.eps}}
 \end{picture}}
 \caption {Multiwavelength unabsorbed spectra for the Vela and Geminga pulsars from the radio to hard $\gamma$-rays  
 compiled from the data obtained with different instruments, as indicated in the plots. Mid-IR fluxes are marked by red circles.
 Down triangles mark upper limits.
Filled squares in the radio range of Geminga  are its radio detections by \citet{Mal1997}. 
}
 \label{fig:mw1}
 \end{figure*}

Besides  the Crab, where the excess  is likely to be  marginal  \citep{sol09}, 
there are only two other  isolated NSs detected  
in the mid-IR. These are  the AXPs 4U 0142+61 \citep{Wang2006}  and 1E 2259+586  \citep{Kaplan2009}. 
Both have been widely considered as radio-silent\footnote{See, however, \citet{Mal2005,Mal2010}.}, 
and belong to a class of magnetars, 
i.e., NSs with extremely strong magnetic fields.   
They have significant mid-IR excesses, which have been interpreted 
as the emission from  X-ray irradiated fall-back discs around the NSs.  
Such  discs can be formed around isolated NSs through a process of supernova fall-back 
\citep{chev1989}.
 
The marginal excess  over the power-law extrapolation of the optical SED   for 
the Crab was suggested to be caused by a nearby PWN knot located in 0\farcs6 from the pulsar. The knot  is much fainter 
than the pulsar in the optical, but  it has a very red spectrum  with a steep flux increase towards the mid-IR,  where 
the knot brightness can be comparable to that of the pulsar \citep{sol09}.  This can result in 
 an overestimation  
of the measured pulsar flux in this range, since the \textit{Spitzer} spatial resolution does not allow 
one to resolve the knot from the pulsar.  

A planet, or a cool low-mass stellar companion 
could also provide an excess. However, 
this is unlikely for our pulsars, since 
no planets or binary companions have been reported from timing observations of Vela and 
Geminga\footnote{Possible planet orbiting Geminga \citep{mattox98} hasn't been confirmed by recent analysis.}. 

Finally, the observed excesses can be produced in magnetospheres of the pulsars.  
The latter possibility has also not been excluded  for the magnetar 
1E 2259+586  \citep{Kaplan2009}.   
Below we discuss  in more detail the possibilities listed above for 
the interpretation of   the Vela 
and Geminga SEDs. We will focus mainly on the Vela pulsar, 
whose mid-IR counterpart candidate is  firmly detected.     
    
  \subsubsection{The fall-back disc}
 
As protoplanetary discs around ordinary pre-main sequence stars  \citep[e.g.,][]{Dull2010} or dust discs 
detected around white dwarfs \citep[e.g.,][]{melis2010, farihi2010}, the fall-back discs around pulsars can lead to emission  
excesses  in  pulsar spectra in the mid-IR and submillimetre bands \citep{Foster1996, Perna2000}. 

To find out whether  the Vela and Geminga spectral excesses can 
be explained by radiation from  hypothetical discs around these objects,  
following \citet{Wang2006}, we consider  passive dust discs  
heated  by  X-ray irradiation  from the pulsars. 
We adopt a model of the irradiated disc emission from 
\citet{Vrt1990}.  It is assumed that there is enough gas to 
provide the pressure support for the dust, but  its contributions to the heating balance due to 
a viscous dissipation  and the emission  are negligible. 
In this model the disc is optically thick and at a fixed disc radius it radiates as blackbody   
with a temperature depending on the radius $\propto R^{-3/7}$ \citep{Vrt1990}.      
The  parameters determining the emission properties of the disc  are  the  inclination angle   
$\zeta$,   albedo $\eta$,  and the internal and external disc radii $R_{in}$ and $R_{ex}$. 
All of  them  cannot  be certainly constrained from the spectral fit to the observed IR SEDs containing only  four points 
and upper limits. One has to select most critical ones to be found from the fit, 
and fix the others at  reasonable values. Albedo  strongly affect the results and 
in our fits  we fixed  it using   a set of  values between 0.5 and 1. We fixed also  possible 
contributions from the pulsars in the IR, assuming that in this range 
their spectra are consistent with a power-law extrapolation from  the optical--UV range.  
The observable disc properties  depend also on the X-ray luminosity of the pulsar and its distance, 
which  are known   for our pulsars from previous studies: $L_X^{Vela}$ = (5.3$\pm$0.5)$\times$10$^{32}$  erg s$^{-1}$ 
\citep{pavlov2001},  $L_X^{Gem}$ = (3$\pm$1)$\times$10$^{31}$  erg s$^{-1}$ \citep{Kar2005}, 
$d^{Vela}$ = 287$^{+19}_{-17}$ pc \citep{Dod2003}, and $d^{Gem}$ = 250$^{+120}_{-60}$ pc \citep{fah07}. 
 
  \begin{figure} 
\begin{center}
\includegraphics[scale=0.35, clip]{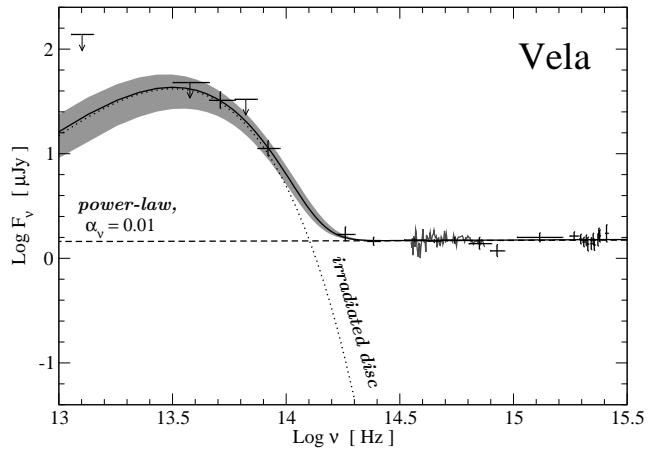}
\end{center}
 \caption {The spectral fit of the dereddened Vela pulsar optical--IR SED by a sum of the fixed power-law component (dashed line), which describe
  contribution from the pulsar itself, and 
  X-ray irradiated passive dust disc model (dotted line). The solid line is the best-fitting result and  
  the hatched region shows 1$\sigma$ uncertainty of the fit.  
  The dotted line is the contribution from the thermally emitting disc with the best-fitting parameters.    
  For the parameters see the text.}
 \label{fig:v-fit}
 \end{figure}    
For Vela the inclination angle can be fixed at $64^{\circ}$, accounting for the viewing angle  of  
the pulsar rotation axis,    estimated  based  on  the PWN geometry \citep{Helfand2001, Ng2008}. 
Within uncertainties this  angle  is  compatible with  that derived from the polarimetric radio observations 
of the pulsar  \citep{John2005}. We assume, that  the disc and pulsar rotation axes   coincide with each other.  
As a result, the irradiated disc model provides a good fit for the Vela SED    
at $\eta$ $<$ 0.85. An example is presented in  Fig.~\ref{fig:v-fit} where we fixed $\eta$ at 0.75. 
In this case $R_{in} = 0.28 \pm 0.05$ R$_{\odot}$,  $R_{ex} = 1.4 \pm 0.4$ R$_{\odot}$,
and the dust temperatures at the inner and external disc boundaries  
$T_{in} = 834 \pm 63$ K, and $T_{ex} = 418 \pm 56$ K, respectively.

The $R_{in}$ value derived from the fit  is  much larger 
than the pulsar light cylinder  radius  $R_{LC}$ $\approx$ 0.6$\times$10$^{-2}$ R$_{\odot}$.  
This allows  Vela to work as the radio pulsar.   At the same time  it is about ten times   
smaller than  that of AXP 4U 0142+61 \citep{Wang2006}, 
suggesting a more compact disc around the Vela pulsar than  the disc around the AXP, 
while the inner disc temperatures for both NSs are similar and consistent  with the sublimation 
temperature of dust $\ga$ 1000 K.  This is not a surprise, since the X-ray luminosity of  
the AXP, 9$\times$10$^{35}$ erg s$^{-1}$, is about three orders of magnitude higher than that of  the Vela pulsar,  
and the AXP dust disc can survive   only at larger distances from the NS.  
Therefore, $R_{in}$ derived from the fit is in agreement  with  the inner disc boundary, that can be estimated 
independently  from the dust destruction temperature, as it has  also been noticed  by \citet{Wang2006} for 
the 4U 0142+61 case.  
We also note, that our $R_{ex}$ value is compatible with a condition 
$R_{ex}$ $\la$ R$_{\odot}$ found to be correct for almost all dust 
discs observed around white dwarfs \citep{farihi2010}. This means
that $R_{ex}$ cannot exceed the Roche radius, within which any rocky object
would be tidally disrupted \citep{melis2010}. 
The Roche radius for NSs, as for white dwarfs, 
is about  R$_{\odot}$\footnote{The exact value depend on the density 
of a rocky object as well as on mass 
and radius of a star \citep{melis2010}.}.

If $R_{in}\gg R_{LC}$, then fall-back discs around isolated pulsars can be swept away by radiation pressure 
of the intensive magneto-dipole radiation of the pulsar \citep{shvarts070}.  
However, it was shown \citep{eksi2005},  
that such a disc is not disrupted by this process, if $R_{in}$ $\la$ $2R_{LC}$ and  
$R_{in}$ $\la$ $100R_{LC}$ in  cases of the orthogonal and aligned magnetic rotators, respectively.  
For Vela our best fit  value $R_{in}$ $\approx$ $45R_{LC}$  lies in the middle of this range, providing  stability 
of the disc against the radiation pressure at the angles  between the magnetic and rotational axes 
of the pulsar $\la$ 10$^{\circ}$. This is in contrast to a value of $\sim$ 70$^{\circ}$ estimated from 
a combination of the pulsar polarimetric observations  \citep{John2005}, and X-ray PWN  geometry 
analysis \citep{Ng2008}. The latter value suggests the disruption of the disc, since 
at this angle we cannot obtain any reasonable 
fit to the Vela SED with $R_{in}$, that is close enough to the $R_{LC}$ value to satisfy  
the stability criterion. Nevertheless, if the Vela pulsar is indeed closer to the orthogonal 
rotator, but not to the aligned one, 
the disc might still survive,  if  we assume that  optically thin gas disc component  exists  
inside  $R_{in}$, as it is  for dust discs around pre-main-sequence 
stars \citep{Dull2010}. This component cannot contribute significantly to the IR emission, 
but may support the stability of the disc against the radiation pressure.

At the same time, the problem of the dust disc structure and stability  in the presence of a strong 
relativistic particle wind from  energetic pulsars is still not  resolved     
\citep[see, e.g.,][ and references therein]{Bryden2006,Jones2007}. The presence of the planetary 
system orbiting PSR B1257+12 \citep{Wols1994} strongly suggests  that its formation must 
involve a preplanetary disc circling the NS. Despite of many efforts \citep{Bryden2006}, 
the most stringed upper limits on the dust disc emission in this system  obtained  
with the \textit{Spitzer}/MIPS, suggests that the dust, if it really exists there, is significantly cooler 
than that derived by us for the Vela pulsar and by  \citet{Wang2006} for AXP 4U 0142+61.  
Therefore, we conclude, that a simplified model of a passive dust 
disc can formally explain the IR excess for Vela, but it is not clear, whether such a disc can 
survive  around this energetic pulsar. 

We note also, that the Vela IR SED can be roughly fitted by 
a pure blackbody model with   $T \ \approx \ 597$ K, and a blackbody radius  of $\approx$ 0.7 R$_{\odot}$. 
The latter  one is too large for a planet or  a cool brown dwarf. This is an additional argument against 
a cool  binary companion of the pulsar as a source of the excess.   

 We have carried out the similar analysis  for Geminga. The  results  
 are very uncertain  due to  larger uncertainties  of its IR fluxes.   
 We can note only that  the fall-back disc around Geminga 
 cannot be excluded by current data. The  inner disc radius seems to be comparable 
 with the Geminga light cylinder radius,  $R_{LC}$ $\approx$ 0.016 R$_{\odot}$, perhaps, 
 suggesting the disc stability, that may lead to  unstable behaviour of Geminga  
 as the radio pulsar \citep{Mal1997}. However, deeper  IR studies   are  necessary 
 to make any definite conclusions.

  \subsubsection{An unresolved PWN structure}  
 \begin{figure}
\begin{center}
\setlength{\unitlength}{1mm}
\resizebox{11.6cm}{!}{
\begin{picture}(120,100)(1,0)
\put (0,36) {\includegraphics[scale=0.37,clip]{fig9.1.eps}}
\put (25,0) {\includegraphics[scale=0.32,clip]{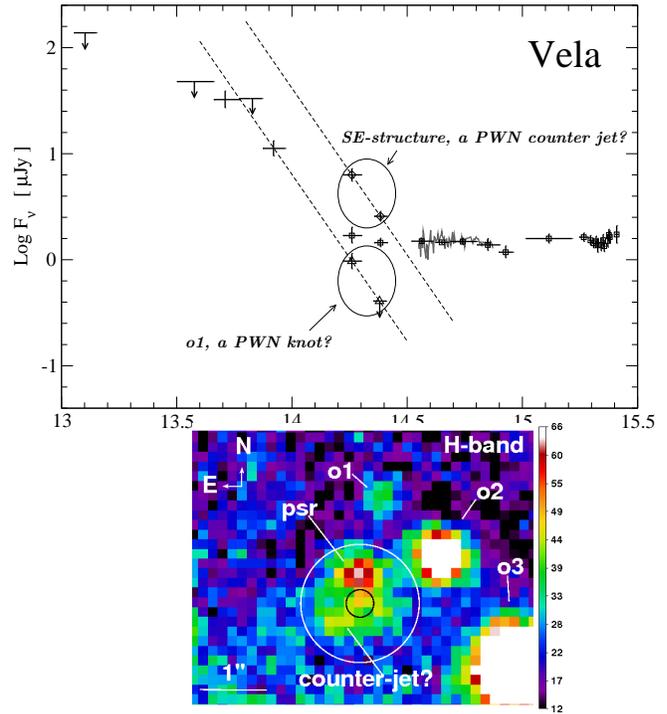}}
 \end{picture}}
  \end{center}
 \caption {The comparison of  the spectral energy distributions for the likely Vela
pulsar counterpart
 and  the  two structures located in  $ \sim$ 2\arcsec~vicinity of the pulsar
and projected on the NW X-ray PWN jet
 ('o1') and the SE counter-jet of the PWN (SE-structure ), shown 
in the  attached fragment of the 
 \textit{H} band image. The  data-points of the  structures are outlined by
ellipses and  notified in the top panel plot, while the rest points belong to the pulsar
counterpart. Dashed lines show  tentative spectral fits  for these structures  by  power
laws with similar spectral indices  
of about 3.1.  As seen, the spectral slope of the pulsar counterpart candidate  in
the IR is compatible with 
the  spectral slopes of the structures. Large and small circles  in the image 
 indicate the FWHM of the PSF and the pulsar position with its $1\sigma$ 
 uncertainty in the mid-IR, respectively   
 }
 \label{fig:fig-o1fit}
 \end{figure}
As in the Crab case, the IR excesses of Vela and Geminga may be produced by unresolved 
nearby structures of their PWNe. 
     
There are  two faint red structures in $ \sim$ 2\arcsec~vicinity of the
Vela pulsar,  which have been detected 
in the near-IR  \citep{shib03}.  They are projected on the X-ray PWN jet and counter-jet, 
and can be barely resolved in the bottom-left panel of Fig.~\ref{fig:vela-ima-1}. For clearness the respective region   is 
enlarged  in the bottom panel of Fig.~\ref{fig:fig-o1fit}.  
A knot-like  object 'o1' is located  $\approx$1\farcs2 NW of the pulsar and detected in the \textit{H} band.  
An extended $\approx$1\farcs5 structure is  located  just SE of the pulsar, 
partially overlapping with it, and  detected in the \textit{J} and \textit{H} bands.  
Both objects are not seen in the optical bands, and the SE structure has showed some 
signs of a short-time  variability in the near-IR  \citep{shib03}. 
The SED of  'o1' with  $F_J$ $\la$ 0.4 $\mu$Jy and $F_H$ = 0.97(18) $\mu$Jy  
can be extrapolated to the mid-IR  using a power-law with a spectral index\footnote{ The index $\alpha_{\nu}$ 
is defined as $F_{\nu}$ $\propto$ $\nu^{-\alpha_{\nu}}$.} $\alpha_{\nu}$ $\approx$ 3 (Fig.~\ref{fig:fig-o1fit}). 
This fact, and the spatial separation of 'o1'  from the pulsar, that is comparable 
to that of a nearby west 
star ('o3', in \citet{shib03} notations) visible in the  mid-IR and optical (cf. Fig.~\ref{fig:vela-ima-1}), 
suggest that  it could be detected  
in the \textit{Spitzer}/IRAC bands at  a similar  brightness level as the pulsar candidate in the wing of its PSF, 
whose mid-IR position and FWHM are marked by  circles in the bottom panel of Fig.~\ref{fig:fig-o1fit}. 
However, we do not detect it  there.  This means, that either 
'o1' has a very peculiar and unrealistic SED with a sharp flux peak in the \textit{H} band, 
or  it is  a time variable object. 
If 'o1' and the SE structure are associated with the Vela PWN, the latter assumption is compatible with 
a strong time variability of the PWN  observed in X-rays \citep{pavlov2001a}.  As seen from  Fig.~\ref{fig:fig-o1fit},  
at the epoch of the near-IR observations  the integral brightness of the SE structure,  
with  $F_J$ = 2.59(12) $\mu$Jy  and  $F_H$ = 4.94(38) $\mu$Jy, was 
higher than the brightness of the pulsar, while 'o1' was fainter.  
The spectral slopes of both structures are  comparable to each other and to the spectral slope 
of the pulsar candidate in the mid-IR.  Since the  SE structure cannot be resolved from the pulsar 
in the \textit{Spitzer} images,  it can considerably contaminate the measured  pulsar fluxes, 
assuming that the emission intensity of this structure in the mid-IR is compatible with the long wavelength 
extrapolation  of its near-IR SED.  This  may be  also supported  by  our astrometry, 
that shows  a marginal  shift of the mid-IR candidate position (black circle in the image of Fig.~\ref{fig:fig-o1fit}) 
from the near-IR counterpart centroid  toward the SE structure.         

We can then speculate,  that the SE structure  can, at least partially, explain the observed excess, 
if this structure  became fainter at the \textit{Spitzer} observation epoch,  as it is for 'o1', 
but saved its steep spectral slope.
In this case, we can expect a variation of the excess intensity with time, 
if the SE structure is indeed associated   with the PWN, that varies with time.  
New \textit{Spitzer}/IRAC observations in 3.6 and 5.8 $\mu$m, and high spatial resolution observations 
in the near-IR \textit{K}$_S$ band  would be useful to verify that.   
We note, that the spectra of 'o1' and  the SE structure are  apparently about 2--3 times steeper 
than the spectrum of the Crab knot with $\alpha_{\nu}$ $\approx$ 1.3 \citep{sol09} and the X-ray 
spectrum of the PWN. The reasons for that are not clear \citep[see, e.g., ][]{shib03},   
and further IR observations can help to understand their real nature.  

The object 'o2' is also not visible in the mid-IR, while it is detected in the optical \textit{RI} and near-IR  
\textit{JH} bands, showing that it is likely to be a distant ($\la$ 10 kpc) main  sequence  background star \citep{shib03}.  
For instance, for M0V star with the observed  colour   \textit{J}$-$\textit{H} $\approx$ 1 at a Galactic $A_V$ $\approx$ 4, we obtain 
d $\approx$ 10 kpc and  the expected  fluxes  at 3.6 and 5.8 $\mu$m of $\la$ 1.5--2.5 $\mu$Jy.  
This is bellow the \textit{Spitzer} observation detection limits and about 20 times  fainter than 
the candidate flux at 5.8 $\mu$m. Therefore, the much brighter candidate  can not be significantly  
contaminated  by 'o2', even if its PSF overlaps with this star.

\subsubsection{Pulsar magnetospheric emission}  
As seen from Fig.~\ref{fig:mw1}, the  nonthermal magnetospheric emission dominates multiwavelength 
spectra of both pulsars from the radio to $\gamma$-rays, except of a narrow range in soft X-rays,  
where the thermal radiation from the surfaces of the NSs is more intensive.  The spectra  are mainly 
of a power-law, but their slopes  are different in different spectral domains,  and there are several 
spectral breaks, e.g., between the  $\gamma$-rays and the hard X-rays,  the soft X-rays and the optical, 
etc. These features are not yet clearly understood, and in this context, a new  
break between the optical and IR does not look very exclusive, and may represent 
an additional component of  the NS magnetospheric emission. 

Except of the Crab, three, of four young PWNe 
so far identified in the optical--IR (PSR B0540$-$69, 3C 58, and G292.0+1.8, \citep{wil08, ser04, shib08, sla08, zhar08, zyuz09}  
show strong  mid-IR excesses with  steep power-law spectral slopes in the IR.  It might be possible, that 
the relativistic particles responsible for these IR excesses are ejected  from magnetospheres of the pulsars 
powering the PWNe. In this case,  the same particles can provide the IR excesses in the magnetospheric emission 
of the pulsars.  
 
However, only the detection of pulsations with pulsar periods  would be a strong evidence 
on the NS magnetospheric origin of  an enhanced emission observed for Vela and tentatively 
for Geminga in the IR.  
Until the time, when timing observations of pulsars in the IR will be possible, imaging observation of other 
suitable rotation powered pulsars in the IR would be useful,  to expand the sample of  NSs detected 
in this range and to verify,  if the IR excess, like that of Vela and Geminga, is a more common feature  in the pulsar 
emission, than a flat optical--IR SED of the Crab.    

\subsection{Mid-IR PWN signatures}  
The inner torus-like PWNe structures around the young Crab, J1124$-$5916, J0205+6449, 
and B0540$-$69 pulsars are  well identified in the optical and mid-IR \citep{sol09, zhar08, zyuz09, shib08, ser04, wil08}.     
The latter three objects  are  even brighter in the IR  than in the optical, and their SEDs   
demonstrate a steep flux increase  toward  longer   wavelengths. In all cases  the optical/IR PWN 
counterparts have approximately the same sizes as in X-rays.      

The Vela and Geminga pulsars are considerably older and less energetic, as compared to the above group, 
and their X-ray PWNe  are fainter as well.   This is likely to be the main reason 
why we do not observe PWNe counterparts  of these pulsars in the optical and IR 
at the same spatial scales as they are seen in X-rays.  
However, we cannot exclude the presence of a hint 
of a counter-jet fragment at a small distance  
from the Vela pulsar. 
As have been discussed above, such poorly resolved small scale structures  can be a reason of 
the observed  IR excesses  in the pulsar SEDs.  The compactness of these structures, 
if they are real,   show  that  the inner PWN evolves with the pulsar age 
and fade  more rapidly  at longer   wavelengths 
than in X-rays.   We note also, that there are still no reports on 
the detection of the inner tours-like 
structure with jets for the Vela PWN in the radio range, while a 10 arcsec
long Geminga tail structure was recently detected with the VLA at 4.5 GHz \citep{pell2011}.
The Geminga radio flux density is compatible with the long wavelengths extension of the tentative
mid-IR excess reported here.     

At the same time, the mid-IR observations allowed us to define more exactly 
the Geminga X-ray PWN shape 
in the immediate vicinity of the pulsar.  Its  forward-jet and the south-east protrusion 
are  likely   not to be  the real PWN structures, but are  X-ray  counterparts of  background 
red stellar object 's1' and 's2' (Fig.~\ref{fig:gem-ima-4}). The same is true for the distant 
blob C, that coincides with a bright IR background object. More careful X-ray spectral 
analysis is necessary to confirm or reject the nature of the 's1' and 's2' objects, that is outside 
the scope of this paper.   

The comparison of the X-ray and 8 $\mu$m images of Vela  shows a good coincidence 
of the west boundary of the extended IR emission of the remnant with the east boundary 
of the large scale X-ray plerion  (Fig.~\ref{fig:vela-ima-3}).  
Further comparison of the mid-IR image with the high spatial resolution radio 
images of the Vela plerion, obtained with the ATCA  by \citet{Dod2003a}, 
likely supports this statement. A  bay  between the north and south plerion  
lobes in the radio image  is filled by a bright protrusion of the extended 
mid-IR emission. The magnetic field lines lie along the bay/protrusion boundary, 
also suggesting that we likely see an interface between the relativistic pulsar wind 
and a dust SNR ejecta provided by the pressure balance between these 
two different parts of the remnant.  No lobes  structures  are visible 
in the optical  \citep[cf.,][]{mign03} and more careful analysis 
is necessary to check this hypothesis.     

To summarize, we conclude, that \textit{Spitzer} mid-IR observations have opened a new spectral window 
for the study of the pulsars. The  detections  of the likely counterparts 
of the Vela and Geminga pulsars, and  PWNe for several Crab-like pulsars reveal  
strong mid-IR excesses in the emission  of these objects, that was rather 
difficult to expect in advance. Further mid-IR studies of these and other 
pulsars  are necessary to increase 
the number of  the objects detected in the IR,
and understand the nature of their emission in this range. 
  
\section*{Acknowledgments} 
This work is based on observations made with the \textit{Spitzer} Space Telescope, 
which is operated by the Jet Propulsion Laboratory, California Institute of Technology under a 
contract with NASA. We are grateful to anonymous referee for useful comments
improving the paper. AAD, DAZ and YAS acknowledge support from the Russian Foundation for Basic 
Research (grants 08-02-00837 and 09-02-12080), Rosnauka
(Grant NSh 3769.2010.2), and the Ministry of  Education and Science of the Russian Federation (Contract No. 11.G34.31.0001). 
SVZ acknowledges support from CONACYT 48493 and PAPIIT IN101506 projects. 
DAZ acknowledges support from St. Petersburg Government grant for young scientists (2.3/04-05/008).


\label{lastpage}

\end{document}